\documentclass[12pt]{article}
\usepackage{latexsym,epsfig,graphicx,epstopdf,amsmath,amssymb,amscd,multirow,chicago,psfrag,paralist,dsfont,url}
\usepackage[titletoc]{appendix}
\usepackage{color}

\usepackage[american]{babel}
\usepackage{dsfont}
\usepackage{algorithmicx}
\usepackage{algorithm}
\usepackage{algpseudocode}
 \usepackage{hyperref}

\newcommand{\thetavec}{{\boldsymbol{\theta}}}

\newcommand{\zvec}{{\boldsymbol{z}}}

\newcommand{\Indfun}{{\mathds{1}}}

\newcommand{\pr}{{\rm Pr}}

\newcommand{\DATA}{{\text{DATA}}}

\newcommand{\xvec}{\boldsymbol{x}}

\newcommand{\lik}{\mathcal{L}}

\newcommand{\tauh}{\tau_{\text{h}}}
\newcommand{\taut}{\tau_{\text{t}}}
\newcommand{\taud}{\tau_{\text{d}}}

\newcommand{\nh}{n_{\text{h}}}
\newcommand{\nt}{n_{\text{t}}}
\algnewcommand\algorithmicassume{\textbf{Assume:}}
\algnewcommand\Assume{\item[\algorithmicassume]}
\algnewcommand\algorithmicrequired{\textbf{Required inputs:}}
\algnewcommand\Required{\item[\algorithmicrequired]}
\algnewcommand\algorithmicreturns{\textbf{Return:}}
\algnewcommand\Returns{\item[\algorithmicreturns]}

\begin{document}

%%%%%%%%%%%%TITLE%%%%%%%%%%%%%%%%%%%%%%%%%%%%%%%%%%%
\title{Planning Reliability Assurance Tests for Autonomous Vehicles}
%%%%%%%%%%%%%%%%%%%%%%%%%%%%%%%%%%%%%%%%%%%%%%%%%%%%%%%%%%%%%%%%%%%%%%%%%%%%%%%%%%%%%%%%%%%%%%%%%%%%%%%%%%%%%%%%%

%\iffalse
\author{
Simin Zheng$^1$, Lu Lu$^2$, Yili Hong$^1$, and Jian Liu$^3$\\[1.5ex]
{\small $^1$Department of Statistics, Virginia Tech, Blacksburg, VA 24061}  \\
{\small $^2$Department of Mathematics \& Statistics, University of South Florida, Tampa, FL 33620}\\
{\small $^3$Department of Systems \& Industrial Engineering, University of Arizona, Tucson, AZ 85721} 	
}
%\fi

%\date{\today}
\date{}

\maketitle
%%%%%%%%%%%%%%%%%%%%%%%%%%%%%%%%%%%%%%%%%%%%%%%%%%%%%%%%%%%%%%%%%%%%%%%%%%%%%%%%%%%%%%%%%%%%%%%%%%%%%%%%%%%%%%%%
\begin{abstract}
Artificial intelligence (AI) technology has become increasingly prevalent and transforms our everyday life. One important application of AI technology is the development of autonomous vehicles (AV). However, the reliability of an AV needs to be carefully demonstrated via an assurance test so that the product can be used with confidence in the field. To plan for an assurance test, one needs to determine how many AVs need to be tested for how many miles and the standard for passing the test. Existing research has made great efforts in developing reliability demonstration tests in the other fields of applications for product development and assessment. However, statistical methods have not been utilized in AV test planning. This paper aims to fill in this gap by developing statistical methods for planning AV reliability assurance tests based on recurrent events data. We explore the relationship between multiple criteria of interest in the context of planning AV reliability assurance tests. Specifically, we develop two test planning strategies based on homogeneous and non-homogeneous Poisson processes while balancing multiple objectives with the Pareto front approach. We also offer recommendations for practical use. The disengagement events data from the California Department of Motor Vehicles AV testing program is used to illustrate the proposed assurance test planning methods.

\end{abstract}

%%%%%%%%%%%%%%%%%%%%%%%%%%%%%%%%%%%%%%%%%%%%%%%%%%%%%%%%%%%%%%%%%%%%%%%%%%%%%%%%%%%%%%%%%%%%%%%%%%%%%%%%%%%%%%%%%%%%%
%\newpage
%\tableofcontents
\newpage

%%%%%%%%%%%%%%%%%%%%%%%%%%%%%%%%%%%%%%%%%%%%%%%%%%%%%%%%%%%%%%%%%%%%%%%%%%%%%%%%%%%%%%%%%%%%%%%%%%%%%%%%%%%%%%%%%%%%%
\section{Introduction}\label{sec:Introduction}
%%%%%%%%%%%%%%%%%%%%%%%%%%%%%%%%%%%%%%%%%%%%%%%%%%%%%%%%%%%%%%%%%%%%%%%%%%%%%%%%%%%%%%%%%%%%%%%%%%%%%%%%%%%%%%%%%%%%%
\subsection{Background and Motivation}\label{subsec:background}

The application of artificial intelligence (AI) technology is growing rapidly and significantly impacting our daily lives. Automation with inherent AI is increasingly emerging in diverse applications. Typical applications of AI technology include fraud protection, automated administrative tasks, autonomous vehicles (AV), facial recognition, and so on. Fueled by big data from advanced computing resources and algorithms, AV plays an important role in the application of AI for improving lifestyle. To ensure that the AVs can be used with confidence, it is necessary to demonstrate their reliability based on statistical methods for assurance test. Traditionally, reliability demonstration tests are commonly used in the product development and assessment process in the fields of industrial engineering, electrical engineering, and health care, to guide the decision on the acceptance of the products based on laboratory data.

Common data types for the reliability analysis include failure time data, recurrent events data, and degradation data. For AV testing, recurrent events data are available. A program of AV testing was launched by the California (CA) Department of Motor Vehicles (DMV) in 2015. Under this program, AV manufacturers are allowed to test AVs on the roads in CA. As part of their agreement, AV manufacturers are required to report (1) annual collision events (CA DMV, \citeyearNP{collision}), (2) mileage information (CA DMV, \citeyearNP{mileage}) as well as (3) annual disengagement events (CA DMV, \citeyearNP{disengagement}), in autonomous mode to the CA DMV. The reported data are accessible to the public for review and assessment. Because of the availability of the recurrent events data for AV testing, this paper focuses on planning reliability assurance tests based on recurrent events data.

Based on these reported data, this paper utilizes the disengagement events and mileage information provided by each manufacturer, as reported at the vehicle identification number (VIN) level, from the CA DMV AV testing program. Disengagement events happen when failures are detected in the technology, communication, sensor, or data reception system. Under these situations, the driver is informed about the autonomous failure by the AV, and is required to take control of the vehicle. Based on the understanding about disengagement events, the recurrent rate of disengagement events can be regarded as a proxy for the reliability of the AV.

Due to the limited research that has been done about the reliability assurance tests for AV using statistical methods, the main goal of this paper is to develop statistical methods for planning AV reliability tests based on recurrent events data. Specifically, to select a best test plan for AV that simultaneously balances multiple objectives, we develop strategies based on homogeneous and non-homogeneous Poisson processes, to investigate the inherent relationships between four test planning criteria including: (1) consumer's risk, (2) producer's risk, (3) the acceptance probability, and (4) the total testing period or the testing period per vehicle. We utilize the Pareto front approach to identify superior test plans based on simultaneously balancing multiple objects. To illustrate the proposed assurance testing plans, we use the data released from the CA DMV AV driving program.

%%%%%%%%%%%%%%%%%%%%%%%%%%%%%%%%%%%%%%%%%%%%%%%%%%%%%%%%%%%%%%%%%%%%%%%%%%%%%%%%%%%%%%%%%%%%%%%%%%%%%%%%%%%%%%%%%%%%%
\subsection{Related Literature and Contribution of This Work}\label{subsec:literature.review}
%%%%%%%%%%%%%%%%%%%%%%%%%%%%%%%%%%%%%%%%%%%%%%%%%%%%%%%%%%%%%%%%%%%%%%%%%%%%%%%%%%%%%%%%%%%%%%%%%%%%%%%%%%%%%%%%%%%%%

As AI systems being more and more popular in a variety of applications, several studies have been done in the field of the reliability or robustness analysis of AI systems.  \shortciteN{Xie2019Talk} discussed the potential opportunities and current challenges about analyzing reliability of AI systems, and pointed out the importance of reliability analysis of AI systems. \shortciteN{AlshemaliKalita2020} presented a comprehensive review of the methods for improving the robustness of the natural language processing in the field of AI. \shortciteN{hong2023statistical} provided statistical perspectives on the reliability of AI systems and introduced a ``SMART'' statistical framework for AI reliability research. Despite the fast emergence of AI systems and their proceed applications, statistical analysis of AI reliability remains in its early stage of development.

Some studies have investigated reliability analysis of AI in AVs. \shortciteN{kalra2016driving} applied the statistical hypothesis testing approach to calculate the number of driving miles that is needed for demonstrating AV reliability. \shortciteN{Merkel2018} applied the software reliability growth models (SRGMs) including Musa-Okumoto model and Gompertz model for estimating and predicting the reliability based on the CA public-road testing data. \shortciteN{monkhouse2020enhanced} created an enhanced vehicle control model that expands the concept of controllability and joint cognition for highly automated tasks. \shortciteN{khastgir2021systems} expanded the systems theoretic process analysis method to identify test scenarios for AV driving systems. \shortciteN{min2022reliability} introduced a statistical framework for modeling and analyzing recurrent events data from AV driving tests using parametric and non-parametric methods, to determine the reliability of the AI system in AVs. \shortciteN{tao2022short} investigated short-term AV maintenance planning, specifically for autonomous trucks, aiming to identify low-risk maintenance decisions. \shortciteN{pauer2022introducing}  introduced a new safety assessment method using a simplified binary integer AV model to optimize the process, with a focus on AV system safety.

This paper focuses more on the aspect of designing statistical assurance test based on homogeneous Poisson process (HPP) and non-homogeneous Poisson process (NHPP) models for analyzing the recurrent events data. There exist several related works in this area. \shortciteN{Hamadaetal2008} introduced the background, general methodologies for modeling repairable systems and recurrent events data. \shortciteN{lu2016multiple} developed a multi-objective decision-making platform for non-repairable systems, based on the binomial demonstration test. \shortciteN{kim2019cost} proposed a reliability demonstration method using an accelerated degradation test within a nonlinear random-coefficients model framework. \shortciteN{wang2019multi} investigated a multi-phase reliability growth test planning approach for repairable products with independent competing failure modes. \shortciteN{hamada2020assurance} considered assurance testing for repairable systems based on both the HPP and NHPP models under a Bayesian framework and also developed an algorithm for finding an assurance test. \shortciteN{wilson2021assurance} developed the assurance approach for the sample size calculation in reliability demonstration testing for binomial and Weibull distributions. While there are established statistical methods available for demonstrating reliability, there is limited research on integrating these methods into the design and test planning for AV reliability.

Several research studies have been conducted using the publicly available CA DMV self-driving data. \shortciteN{dixit2016autonomous} and \shortciteN{favaro2017examining} presented comprehensive analysis for accidents events  data based on the public CA AV testing data. \shortciteN{zhao2019assessing} proposed a new Bayesian method to access the safety and reliability of AVs and studied the trend of disengagements by applying SRGMs to the CA public road testing data. \shortciteN{Boggsetal2020} conducted an exploratory analysis of AV collision events data using text analytics and hierarchical Bayesian heterogeneity-based approach. \shortciteN{sinha2021crash} provided a general introduction and visualization of the disengagement events data on public roads in CA from 2014 to 2019. Although there have been many studies examining the CA DMV public testing data, only a few have employed a thorough statistical method for planning reliability tests using this public dataset.

In addition, we use the Pareto front optimization approach to make better decisions based on multi-objectives for the assurance test of AVs. \shortciteN{rachmawati2009multiobjective} proposed a selection scheme, which allows a multi-objective evolutionary algorithm to generate a non-dominated set with adjustable concentration surrounding the optimal tradeoff region. \shortciteN{lu2011optimization} advanced the Pareto front approach by developing a structured two-stage decision-making process to efficiently examine and select optimal designs. \shortciteN{khorram2014numerical} introduced a numerical approach to construct an approximation of the Pareto front in multi-objective optimization problems.  \shortciteN{hua2021survey} proposed a comprehensive review for the research on multi-objective optimization problems with irregular Pareto fronts.

In summary, while there have been numerous studies examining various aspects of reliability demonstration testing, statistical methods have not been employed in planning AV tests. The contribution of this work is that we establish a framework for demonstration of AI reliability based on publicly available CA DMV test dataset.

%%%%%%%%%%%%%%%%%%%%%%%%%%%%%%%%%%%%%%%%%%%%%%%%%%%%%%%%%%%%%%%%%%%%%%%%%%%%%%%%%%%%%%%%%%%%%%%%%%%%%%%%%%%%%%%%%%%%%
\subsection{Overview}\label{subsec:overview}
%%%%%%%%%%%%%%%%%%%%%%%%%%%%%%%%%%%%%%%%%%%%%%%%%%%%%%%%%%%%%%%%%%%%%%%%%%%%%%%%%%%%%%%%%%%%%%%%%%%%%%%%%%%%%%%%%%%%%
The rest of the paper is organized as follows. Section~\ref{sec:data.model} introduces the data notation and statistical models for recurrent events data, along with a general background on the Bayesian method. Section~\ref{sec:assurance.test.framework} introduces the assurance test framework, including three primary risk types, which are often considered in assurance tests. Mathematical details and algorithms for computing these risks will also be provided. Section~\ref{sec:HPP.main} explores the relationship and trade-off between multiple criteria under the HPP model for recurrent events data. Pareto front approach will be used to select optimal test plans based on considering different testing priorities. Section~\ref{sec:NHPP.main} extends the method for NHPP model. Section~\ref{sec:conclusion} contains some concluding remarks and potential areas for future research.

%%%%%%%%%%%%%%%%%%%%%%%%%%%%%%%%%%%%%%%%%%%%%%%%%%%%%%%%%%%%%%%%%%%%%%%%%%%%%%
\section{Data and Statistical Models}\label{sec:data.model}
%%%%%%%%%%%%%%%%%%%%%%%%%%%%%%%%%%%%%%%%%%%%%%%%%%%%%%%%%%%%%%%%%%%%%%%%%%%%%%%%%%%%%%%%%%%%%%%%%%%%%%%%%%%%%%%%%%%%%
\subsection{Notation for data}\label{subsec:notation}
%%%%%%%%%%%%%%%%%%%%%%%%%%%%%%%%%%%%%%%%%%%%%%%%%%%%%%%%%%%%%%%%%%%%%%%%%%%%%%%%%%%%%%%%%%%%%%%%%%%%%%%%%%%%%%%%%%%%%

To design a reliability assurance test for AVs, first, we define various time periods. The historical data period refers to the time window during which historical data were collected. The historical data were then used to derive the posterior distribution of model parameters for the subsequent test planning. We denote the historical data period as $[0, \tauh]$, where $\tauh\geq 0$, with the sample size of the historical data denoted by $\nh$. Note that when $\tauh=0$, it suggests there is no historical data available. Then the testing period is the time interval we perform the assurance test, which is denoted as $(\tauh, \tauh+\taut]$, where $\taut>0$, with the sample size of the assurance test denoted as $\nt$. Lastly, the demonstration period is the time window where the reliability metric will be evaluated at the end of the duration, and it can be defined as $(\tauh, \tauh+\taud]$, where $\taud \geq 0$, and $\taud \geq \taut$. Note that while the demonstration period is often anticipated to be substantially longer than the testing period, these two time periods usually overlap.

This paper uses historical data from December 1, 2017, to November 30, 2019, which is a two-year study period, thus $\tauh=2\times 365= 730$ days.
More specifically, the disengagement events data are structured as recurrent events data, reported by each manufacturer at the VIN level. As for the mileage information, the public AV testing data reports only monthly mileage, so daily mileage is calculated by dividing the monthly mileage information by the number of days in that month. This assumes a constant daily mileage for each vehicle throughout the month as in \shortciteN{min2022reliability}.

Then for the historical data, the time to events during the historical data period are denoted as $t_{ij}$ for the $i$th test unit at the $j$th recurrent event, where $i=1, \dots, \nh$ and $j=1, \dots, n_i$. We use $n_i=0$ to denote that no event was observed for unit $i$ in the historical data period $[0, \tauh]$. Let $x_i(t)$ denotes the mileage driven by unit $i$ at time $t$ (in a day), where $0<t \leq \tauh$. The unit of $x_i(t)$ is k-miles. We also define $\xvec_i(t)=\{x_i(s): 0<s\leq t \}$ as the historical daily mileage records driven by unit $i$ for a given interval.

%%%%%%%%%%%%%%%%%%%%%%%%%%%%%%%%%%%%%%%%%%%%%%%%%%%%%%%%%%%%%%%%%%%%%%%%%%%%%%%%%%%%%%%%%%%%%%%%%%%%%%%%%%%%%%%%%%%%%
\subsection{Statistical Models for Recurrent Events Data}\label{subsec:stat.models}
%%%%%%%%%%%%%%%%%%%%%%%%%%%%%%%%%%%%%%%%%%%%%%%%%%%%%%%%%%%%%%%%%%%%%%%%%%%%%%%%%%%%%%%%%%%%%%%%%%%%%%%%%%%%%%%%%%%%%

Recurrence events data are often modeled with HPP or NHPP models. Considering the HPP is a special case of NHPP, we begin with introducing the more general NHPP model and then discuss the more specific HPP model. Specifically, if we assume there is no reliability growth during the testing and demonstration periods, then the event intensity is constant, which is the case of HPP. When we assume there is reliability growth during the test period, that is when we have updated the system over time, it is the case of NHPP.

Under NHPP, the number of events occurring in the time window $(0,t]$ is assumed to follow a Poisson distribution with a non-constant intensity function $\lambda(t)$, for $t>0$. More specifically, the event intensity function for unit $i$ at time $t$ is modeled as:
\begin{align}\label{eqn:lambda.t.general}
 \lambda_{i} [t;\xvec_i(t),\thetavec] = \lambda_{0} (t;\thetavec) g[x_{i}(t)],
 \end{align}
where $\lambda_{0} (t;\thetavec)$ denotes a non-constant baseline intensity function (BIF) which varies over time and the parameter $\thetavec$ represents the vector of unknown parameters in the model. Also, $g(\cdot)$ can be substituted with a specific form based on the particular analysis, and $x_{i}(t)$ is the mileage for unit $i$ at time $t$, as introduced in Section~\ref{subsec:notation}.   Following \shortciteN{min2022reliability}, we use $g[x_{i}(t)]=x_{i}(t)$ in this paper, which means the intensity is proportional to the mileage driven. However, our method can also be extended to other functional forms of $g(\cdot)$. In summary, $\lambda_{i} [t;\xvec_i(t),\thetavec]$ is the mileage-adjusted event intensity since $g[x_{i}(t)]$ is the mileage effect function.

Additionally, the cumulative baseline intensity function (CBIF) is given by:
\begin{align}\label{eqn:baseline.CBIF}
\Lambda_{0} (t;\thetavec)  = \int_{0}^{t}\lambda_{0} (s;\thetavec)ds,
 \end{align}
where $\Lambda_{0}(t;\thetavec)$ is a non-decreasing function of time $t$ and $\Lambda_{0}(0;\thetavec)=0$. The CBIF can be interpreted as the expected number of failure events occurs in the time period $(0,t]$. Then, the cumulative intensity function (CIF) for unit $i$ is calculated as:
\begin{align}\label{eqn:CIF}
\Lambda_{i} [t;\xvec_i(t),\thetavec] = \int_{0}^{t} \lambda_{0} (s;\thetavec) g[x_{i}(s)]ds.
\end{align}

As a special case of NHPP, HPP assumes the event intensity function for unit $i$ at time $t$ to be:
\begin{align}\label{eqn:lambda.t.special.case}
  \lambda_{i} [t;\xvec_i(t),\thetavec] = \lambda_{0} (\thetavec)g(x_{i}),
 \end{align}
where $\lambda_{0}(\thetavec)$ denotes the BIF which does not vary over time. Hence, it is simplified as $\lambda_{0}(\thetavec)=\lambda_{0}$ for the HPP. The BIF can take different forms when using different parametric models, such as Musa-Okumoto, Gompertz, or Weibull model, see for example, \shortciteN{min2022reliability}. Here, $\lambda_{0}$ represents the rate of failure events per k-miles. In addition, assume that the mileage effect function remains constant over time for each unit $i$, and hence denoted as $g[x_{i}(t)]=g(x_{i})=x$.

\subsection{Data and Bayesian Analysis}\label{subsec:data.Bayesian}

This paper designs the reliability assurance test plans for AVs using the posterior distribution derived from the CA DMV public driving test data from December 1, 2017, to November 30, 2019. First, we collect online public data with a focus on annual disengagement events and mileage. Then, several data cleaning steps are required to derive the daily mileage information and prepare the final format of the disengagement events data for each VIN. This two-year dataset, after data cleaning, can be considered the original historical dataset.

Using Bayesian analysis principles, we combine two-year historical data obtained from the CA DMV AV test program with the user-specified priors $p(\thetavec)$ by applying Bayes' theorem to derive the posterior distribution $\pi(\thetavec\vert \DATA)$, with a primary focus on Waymo manufacturer due to its extensive on-road testing during the study period. More specifically, in Bayesian analysis, to derive $\pi(\thetavec\vert \DATA)$, we first need to derive the likelihood function $\lik(\thetavec \vert \DATA)$, which is a function of  $\thetavec$. The likelihood function is as follows:
\begin{align}\label{eqn:likelihood}
\lik(\thetavec \vert \DATA) &= \prod_{i=1}^{\nh} \left\{\prod_{j=1}^{n_{i}}\lambda_{i}[t_{ij};\xvec_{i}(t_{ij}),\thetavec]\right\} \times \exp\{ -\Lambda_{i}[\tauh,\xvec_{i}(\tauh),\thetavec]\},
\end{align}
where $\prod_{j=1}^{0}(\cdot)=1$ for any unit without any observed event. The event intensity function and CIF are demonstrated in \eqref{eqn:lambda.t.general} and \eqref{eqn:CIF} respectively for the NHPP. While the event intensity function is defined in \eqref{eqn:lambda.t.special.case} for the HPP.

Then, to obtain the posterior distribution of $\thetavec$, we need to apply the Bayes' theorem, which is:
\begin{align}\label{eqn:posterior}
\pi(\thetavec \vert \DATA) &= \frac{\lik(\thetavec \vert \DATA )p(\thetavec )}{\int_{\thetavec}\lik(\thetavec \vert \DATA)p(\thetavec )d\thetavec} \propto \lik(\thetavec \vert \DATA )p(\thetavec),
\end{align}
where $p(\thetavec)$ is the user-specified prior distribution for $\thetavec$.

In this paper, we use the normal priors, where the priors are relatively flat to obtain the posterior distribution $\pi(\thetavec\vert \DATA)$. This can be regarded as our input dataset for the subsequent development of the test planning for AVs under both the HPP and NHPP models. More specifically, we first collect the public AV test data from CA DMV website from 2017 to 2019. The two-year historical public data consist of recurrent events including (1) manufacturer information, (2) vehicle identification number (VIN), (3) disengagement event dates, and (4) annual mileage details for each recorded VIN. Following \eqref{eqn:likelihood} and \eqref{eqn:posterior} with the normal prior $p(\thetavec)$, we can derive the corresponding posterior distribution $\pi(\thetavec\vert \DATA)$ structured with $n_{\text{Post}}$ samples for the Weibull reliability growth model for the three unknown parameters $\theta_{1}$, $\theta_{2}$ and $\theta_{3}$. We used $n_{\text{Post}}=1001$ in our analysis.

%%%%%%%%%%%%%%%%%%%%%%%%%%%%%%%%%%%%%%%%%%%%%%%%%%%%%%%%%%%%%%%%%%%%%%%%%%%%%%%%%%%%%%%%%%%%%%%%%%%%%%%%%%%%%%%%%%%%%
\subsection{Reliability Metrics}\label{subsec:reliability.metrics}
%%%%%%%%%%%%%%%%%%%%%%%%%%%%%%%%%%%%%%%%%%%%%%%%%%%%%%%%%%%%%%%%%%%%%%%%%%%%%%%%%%%%%%%%%%%%%%%%%%%%%%%%%%%%%%%%%%%%%

For recurrent events data, we use the average intensity as the reliability metric, which is defined as
\begin{align}\label{eqn:reliability_metric}
m(s, t)=\frac{\Lambda[t; \xvec(t),\thetavec]-\Lambda[s; \xvec(s),\thetavec]}{t-s}
=\frac{\Lambda(t)-\Lambda(s)}{t-s},
\end{align}
for a unit with cumulative intensity $\Lambda[t; \xvec(t),\thetavec]$ and mileage history $\xvec(t)$. Note that we will use $\Lambda(t)$ and $\Lambda(s)$ just for the purpose of notation simplicity. First, for the NHPP, the average event intensity can be calculated using \eqref{eqn:reliability_metric}. More specifically, for the demonstration period, the average intensity is $m(\tauh, \tauh+\taud)$. To simplify notation, we use $m_{\taud}=m(\tauh, \tauh+\taud)$. In addition, as for the testing period, the average intensity is $m(\tauh, \tauh+\taut)$. Similarly, we will use $m_{\taut}=m(\tauh, \tauh+\taut)$. Then, as a special case of the NHPP, in the HPP model, the average intensity is denoted as $m(s, t)=m$, which is constant over time.

In general, for both the HPP and NHPP models, let $m(s,t)$ denote the actual average failure intensity during the time interval of interest. Let $m_1$ and $m_0$ represent the highest average failure event intensity that could be accepted by the consumers and the lowest average failure event intensity is acceptable for the producers, respectively, where $m_0 \le m_1$. The region $m(s,t) \in (m_0,m_1)$ can be considered as indifference region.

%%%%%%%%%%%%%%%%%%%%%%%%%%%%%%%%%%%%%%%%%%%%%%%%%%%%%%%%%%%%%%%%%%%%%%%%%%%%%%%%%%%%%%%%%%%%%%%%%%%%%%%%%%%%%%%%%%%%%
\section{Reliability Assurance Test Framework}\label{sec:assurance.test.framework}
%%%%%%%%%%%%%%%%%%%%%%%%%%%%%%%%%%%%%%%%%%%%%%%%%%%%%%%%%%%%%%%%%%%%%%%%%%%%%%%%%%%%%%%%%%%%%%%%%%%%%%%%%%%%%%%%%%%%%

%%%%%%%%%%%%%%%%%%%%%%%%%%%%%%%%%%%%%%%%%%%%%%%%%%%%%%%%%%%%%%%%%%%%%%%%%%%%%%%%%%%%%%%%%%%%%%%%%%%%%%%%%%%%%%%%%%%%%
\subsection{Risks in Reliability Assurance Tests}\label{subsec:risks}
%%%%%%%%%%%%%%%%%%%%%%%%%%%%%%%%%%%%%%%%%%%%%%%%%%%%%%%%%%%%%%%%%%%%%%%%%%%%%%%%%%%%%%%%%%%%%%%%%%%%%%%%%%%%%%%%%%%%%
This paper considers the Poisson process assurance test, where a sample of the vehicle units is tested to observe the number of failure events given a specific test duration. Generally, when determining the parameters of a test plan, three types of risks are typically taken into account. These include the consumer's risk (CR), the producer's risk (PR), and the acceptance probability (AP). The \text{CR} is defined as the probability of a product passing the test even though its reliability does not meet the criteria, while the \text{PR} refers to the probability of failing a test, even if the unit's reliability is considered sufficient. The \text{AP}, is the probability of accepting the unit given a successful test.

Bayesian approaches allow researchers to incorporate background knowledge into their analysis. From the Bayesian perspective, these three types of risks can be calculated using the corresponding posterior probabilities, known as the posterior risk criteria. In the following sections, we provide more details for calculating the posterior risk criteria under the Bayesian framework based on the HPP and NHPP models.

%%%%%%%%%%%%%%%%%%%%%%%%%%%%%%%%%%%%%%%%%%%%%%%%%%%%%%%%%%%%%%%%%%%%%
\subsection{Risks Under the HPP Model}\label{subsec:risks.HPP}
%%%%%%%%%%%%%%%%%%%%%%%%%%%%%%%%%%%%%%%%%%%%%%%%%%%%%%%%%%%%%%%%%%%%
Since we use the average intensity as the reliability metric under the HPP model, our primary goal is to demonstrate the average intensity does not exceed the required level of confidence. To choose a test plan, we need to determine the desired planning values with a set of parameters $(\nt,\taut,c)$, where $\nt$ is the number of test units, $\taut$ is the test duration per vehicle, and $c$ is the maximum allowable failures (i.e., disengagement events) to pass the test, i.e., the product is deemed to have met the reliability requirement if the observed event counts $y \le c$. Note that under the HPP model, since the average failure intensity is constant, any combination for $\nt$ and $\taut$ satisfying the total test vehicle days $\tau=n_t\taut$ provides an acceptable test plan.

First, since the average failure intensity is constant over time under the HPP model, under the Bayesian framework, the posterior consumer's risk (PCR) can be calculated as,
\begin{align}\label{eqn:pcr.hpp}
\text{PCR}
&=\Pr(m\geq m_{1} \vert \text{Test is Passed} )
=\Pr(m\geq m_{1} \vert y \le c ) \\\nonumber
&=\frac{\int_{0}^{\infty}\Pr(y\leq c \vert m)\pi(m)\Indfun(m\geq m_1)dm}{\int_{0}^{\infty}\Pr(y\leq c \vert m)\pi(m)dm} =\frac{\int_{0}^{\infty}\left[\sum_{y=0}^{c}h(y; m\tau)\right]\Indfun(m\geq m_1)\pi(m)dm}{\int_{0}^{\infty}\left[\sum_{y=0}^{c}h(y; m\tau)\right] \pi(m)dm}.
\end{align}

When historical data (e.g., two-year CA DMV test data) were used to elicit the prior distribution of $m$, which can be denoted as $\pi(m)$, we can use the posterior distribution of $m$ given the historical data, i.e., $\pi(m \vert \DATA)$, to replace $\pi(m)$ for estimating posterior risk criteria. Hence $\pi(m)$ in \eqref{eqn:pcr.hpp} is the pre-posterior of the failure intensity, which uses the derived posterior distribution as the prior in Bayesian analysis (e.g., \shortciteNP{hong2015bayesian}). Then, $m=\lambda_{0}(\thetavec)g(x_{i})=\lambda_{0}(\thetavec)x$, where $\lambda_{0}(\thetavec)$ is the BIF for the HPP model, and it is a constant function with respect to the parameter vectors $\thetavec$ over two-year historical period as discussed in Section~\ref{subsec:stat.models}. In addition, $\Indfun(\cdot)$ is the indicator function. Note that $h(y; m\tau)$ is the probability mass function (pmf) of a Poisson distribution, which is given in the form:
\begin{align}\label{eqn:Poisson.pmf}
h(y; m\tau)=(m\tau)^{y}\exp{(-m\tau)}/(y!),
\end{align}
for $y=0, 1, \dots$, and $0< m\tau<\infty$.

Similarly, the posterior producer's risk (PPR) can be calculated as
\begin{align}\label{eqn:ppr.hpp}
\text{PPR}
&=\Pr(m \leq m_{0} \vert \text{Test is Failed})=\Pr(m \leq m_{0} \vert y>c) \\\nonumber
%&=\int_{0}^{m_0}p(m \vert y>c)dm \\
&=\frac{\int_{0}^{\infty}\Pr(y>c \vert m)\pi(m)\Indfun(m\leq m_{0})dm }{\int_{0}^{\infty}\Pr(y>c \vert m)\pi(m)dm} =\frac{\int_{0}^{\infty}\left[1-\sum_{y=0}^{c}h(y; m\tau)\right]\Indfun(m\leq m_{0})\pi(m)dm}{\int_{0}^{\infty}\left[1-\sum_{y=0}^{c}h(y; m\tau) \right] \pi(m)dm}.
\end{align}
The acceptance probability (AP), i.e., the probability of passing the test, can be calculated as
\begin{align}\label{eqn:ap.hpp}
\text{AP}
&=\Pr(\text{Test is Passed} ) =\Pr( y \le c ) \\\nonumber
&=\int_{0}^{\infty}\Pr(y\leq c \vert m)\pi(m)dm =\int_{0}^{\infty}\left[\sum_{y=0}^{c}h(y; m\tau)\right] \pi(m)dm.
\end{align}

%%%%%%%%%%%%%%%%%%%%%%%%%%%%%%%%%%%%%%%%%%%%%%%%%%%%%
\subsection{Risks Under the NHPP Model}\label{subsec:risks.NHPP}
%%%%%%%%%%%%%%%%%%%%%%%%%%%%%%%%%%%%%%%%%%%%%%%%%%%

Under the NHPP model, since the failure intensity varies over time, we use the average failure intensity as the reliability metric for characterizing the AV performance. Suppose our goal is to demonstrate the reliability performance over the demonstration period $(\tauh,\tauh+\taud)$ specified by the test objective. We use $(\nt,\taut,c)$ to represent the test plan. When $\nt=1$, the test is a single vehicle test, and when $\nt>1$, the test is a multiple vehicle test. Specifically, if the vehicles with sample size $\nt$ participate in the planned assurance test for $\taut$ days and we observe no more than $c$ failures, then the vehicles will successfully pass the test.

Then the PCR can be calculated as,
\begin{align}\label{eqn:pcr.nhpp}
\text{PCR}
&=\Pr(m_{\taud}\geq m_1 \vert \text{Test is Passed})= \Pr(m_{\taud}\geq m_1 \vert y \leq c) \\\nonumber
 &= A^{-1}\int_{\Theta}\Pr\left(y \leq c \vert \thetavec \right)\pi(\thetavec) \Indfun(\Lambda(\tauh+\taud)-\Lambda(\tauh)\geq m_1 \taud)d\thetavec\\\nonumber
 &= A^{-1}\int_{\Theta}\left[\sum_{y=0}^{c} h(y;\nt m_{\taut} \taut)\right]\Indfun(\Lambda(\tauh+\taud)-\Lambda(\tauh)\geq m_1 \taud)\pi(\thetavec)d\thetavec,
\end{align}
where
\begin{align*}
A & =\int_{\Theta} \Pr\left(y \leq c \vert \thetavec  \right) \pi(\thetavec)d\thetavec =\int_{\Theta}\left[\sum_{y=0}^{c} h(y;\nt m_{\taut} \taut)\right] \pi(\thetavec)d\thetavec.
\end{align*}
Here, $h(y;\nt m_{\taut} \taut)$ is the pmf of a Poisson distribution, which is defined as below,
\begin{align}\label{eqn:Poisson.pmf.nhpp}
h(y; \nt m_{\taut} \taut)=(\nt m_{\taut} \taut)^{y}\exp{(-\nt m_{\taut} \taut)}/(y!),
\end{align}
for $y=0, 1, \dots$, and $0< \nt m_{\taut} \taut<\infty$.

Similarly, the PPR can be calculated by,
\begin{align}\label{eqn:ppr.nhpp}
\text{PPR}
 & = \Pr(m_{\taud}\leq m_0 \vert \text{Test is Failed})  = \Pr(m_{\taud}\leq m_0 \vert y>c) \\\nonumber
 &= B^{-1}\int_{\Theta}\Pr\left(y > c \vert \thetavec \right)\pi(\thetavec) \Indfun(\Lambda(\tauh+\taud)-\Lambda(\tauh)\leq m_0 \taud)d\thetavec  \\\nonumber
 &= B^{-1}\int_{\Theta}\left[1-\sum_{y=0}^{c} h(y;\nt m_{\taut} \taut)\right]\Indfun(\Lambda(\tauh+\taud)-\Lambda(\tauh)\leq m_0 \taud)\pi(\thetavec)d\thetavec,
\end{align}
where
\begin{align*}
B & =\int_{\Theta} \Pr\left(y > c \vert \thetavec  \right) \pi(\thetavec)d\thetavec =\int_{\Theta}\left[1-\sum_{y=0}^{c} h(y;\nt m_{\taut} \taut)\right] \pi(\thetavec)d\thetavec.
\end{align*}
Then, the AP is obtained by,
\begin{align}\label{eqn:ap.nhpp}
\text{AP}
&=\Pr(\text{Test is Passed} )=\Pr( y \le c ) \\\nonumber
&=\int_{\Theta} \Pr\left(y \leq c \vert \thetavec  \right) \pi(\thetavec)d\thetavec =\int_{\Theta}\left[\sum_{y=0}^{c} h(y;\nt m_{\taut} \taut)\right] \pi(\thetavec)d\thetavec.
\end{align}

%%%%%%%%%%%%%%%%%%%%%%%%%%%%%%%%%%%%%%%%%%%%%%%%%%%%%%%%%%%%%%%%%%%%%%%%%%%%%%%%%%%%%%%%%%%%%%%%%%%%%%%%%%%%%%%%%%%%%
\subsection{Algorithm for Computing the Risks for Assurance Tests}\label{subsec:algorithm}
%%%%%%%%%%%%%%%%%%%%%%%%%%%%%%%%%%%%%%%%%%%%%%%%%%%%%%%%%%%%%%%%%%%%%%%%%%%%%%%%%%%%%%%%%%%%%%%%%%%%%%%%%%%%%%%%%%%%%
Considering there is usually no closed-form expression of the Bayesian posterior risks, we develop numeric algorithms to compute the risks associated with assurance tests described in Sections~\ref{subsec:risks.HPP} and~\ref{subsec:risks.NHPP}. In particular, Algorithms~\ref{alg:compute.hpp.risks} and \ref{alg:compute.nhpp.risks} give the details on how to compute the related risks under the HPP model and the NHPP model, respectively.

\begin{algorithm}%[H]
\caption{An algorithm for computing posterior risks under the HPP model}\label{alg:compute.hpp.risks}
\begin{algorithmic}
\Assume We have $M$ draws from $\pi(m)$, for a large $M$. Suppose the $j$th draw is denoted by $m^{(j)}$, where $j=1,2, \ldots, M$. And we consider $\tauh=365\times 2 =730$ days.
\Required (1) $\pi(m)$, $n_{\text{Post}}$ samples for $\thetavec$, (2) $\tau$, (3) $c$, (4) $m_1$, (5) $m_0$, (6) the daily mileage for field usage, $x_{\text{d}}$ and (7) the daily mileage driven for testing, $x_{\text{t}}$.

1. Compute $\lambda_0^{(j)}$ for the $j$th draw from $\pi(m)$, where $j=1, \ldots, M$.

2. Calculate AP using Monte Carlo method as AP $\approx \frac{1}{M}\sum_{j=1}^{M}\left[\sum_{y=0}^{c} h(y; m^{(j)}\tau)\right]$, where $h(y; m^{(j)}\tau)$ is based on \eqref{eqn:Poisson.pmf} and $m^{(j)}=x_{\text{t}}\times\lambda_0^{(j)}$.

3. Use Monte Carlo integration to estimate PPR and PCR, and apply the following conditional statements.

\If{ $\sum_{j=1}^{M} \pr(y>c) =\sum_{j=1}^{M}\left[1-\pr(y \le c)\right]=0$ }
   \State $\text{PPR}=0$.
\Else{ PPR $\approx \left\{\sum_{j=1}^{M}\left[1-\sum_{y=0}^{c}h(y; m^{(j)}\tau)\right]\times \Indfun(x_{\text{d}}\times\lambda_0^{(j)}\leq m_{0})\right\}/C$, where $C=\left\{\sum_{j=1}^{M}\left[1-\sum_{y=0}^{c}h(y; m^{(j)}\tau)\right]\right\}$}.
\EndIf
\If{ $\sum_{j=1}^{M} \Indfun(x_{\taud}\times\lambda_0^{(j)}\geq m_1) =0$ }
   \State $\text{PCR}=0$.
\Else{PCR $\approx \left\{\sum_{j=1}^{M}\left[\sum_{y=0}^{c}h(y; m^{(j)}\tau)\right]\times \Indfun(x_{\text{d}}\times\lambda_0^{(j)}\geq m_1)\right\}/ C_1$ where $C_1=\left\{\sum_{j=1}^{M}\left[\sum_{y=0}^{c}h(y; m^{(j)}\tau)\right]\right\}$}.
\EndIf
\Returns The PCR, PPR, AP, and $\tau=\nt\taut$.
\end{algorithmic}
\end{algorithm}

\begin{algorithm}%[H]
\caption{An algorithm for computing posterior risks under the NHPP model}\label{alg:compute.nhpp.risks}
\begin{algorithmic}
\Assume Consider $\tauh=365\times 2 =730$ days.

\Required (1) $\pi(m)$, (2) $\taut$, (3) $c$, (4) $m_1$, (5) $m_0$, (6) $x_{\text{d}}$, (7) $x_{\text{t}}$, (8) $\taud$, and (9) $\nt$.

1. Compute $\Lambda_{0}^{(j)}(\tauh;\thetavec)$, $\Lambda_{0}^{(j)}(\tauh+\taut;\thetavec)$ and $\Lambda_{0}^{(j)}(\tauh+\taud;\thetavec)$ for the $j$th draw from $\pi(m)$, where $j=1, \ldots, M$ based on  \eqref{eqn:baseline.CBIF}.

2. Calculate AP using Monte Carlo method as AP $\approx \frac{1}{M}\sum_{j=1}^{M}\left[\sum_{y=0}^{c} h(y;\nt m_{\taut}^{(j)} \taut)\right]$, where $h(y; \nt m_{\taut}^{(j)}\taut)$ is calculated based on \eqref{eqn:Poisson.pmf.nhpp}, and $m_{\taut}^{(j)}=x_{\text{t}}\times\left(\Lambda_{0}^{(j)}(\tauh+\taut;\thetavec)-\Lambda_{0}^{(j)}(\tauh;\thetavec)\right)$ based on \eqref{eqn:baseline.CBIF} and \eqref{eqn:reliability_metric}.

3. Use Monte Carlo integration to estimate PPR and PCR based on the following conditional statements.

\If{ $\sum_{j=1}^{M} \pr(y>c) =\sum_{j=1}^{M}\left[1-\pr(y \le c)\right]=0$ }
   \State $\text{PPR}=0$.
\Else{ PPR $\approx \left\{\sum_{j=1}^{M}\left[1-\sum_{y=0}^{c} h(y;\nt m_{\taut}^{(j)} \taut)\right]\Indfun(\Lambda^{(j)}(\tauh+\taud)-\Lambda^{(j)}(\tauh)\leq m_0 \taud)\right\}/D$, where $D=\left\{\sum_{j=1}^{M}\left[1-\sum_{y=0}^{c} h(y;\nt m_{\taut}^{(j)} \taut)\right]\right\}$.}
\EndIf
\If{ $\sum_{j=1}^{M} \Indfun(m_{\taud}^{(j)}\geq m_1) =0$ }
   \State $\text{PCR}=0$.
\Else{ PCR $\approx \left\{\sum_{j=1}^{M}\left[\sum_{y=0}^{c} h(y;\nt m_{\taut}^{(j)} \taut)\right]\Indfun(\Lambda^{(j)}(\tauh+\taud)-\Lambda^{(j)}(\tauh)\geq m_1 \taud)\right\}/D_1$, where $\Lambda^{(j)}(\tauh+\taud;\thetavec)=x_{\text{d}}\times\Lambda_{0}^{(j)}(\tauh+\taud;\thetavec)$, $\Lambda^{(j)}(\tauh;\thetavec)=x_{\text{d}}\times\Lambda_{0}^{(j)}(\tauh;\thetavec)$, and $D_1=\left\{\sum_{j=1}^{M}\left[1-\sum_{y=0}^{c} h(y;\nt m_{\taut}^{(j)} \taut)\right]\right\}$.}
\EndIf
\Returns The PCR, PPR, AP, and $\taut$.
\end{algorithmic}
\end{algorithm}

%%%%%%%%%%%%%%%%%%%%%%%%%%%%%%%%%%%%%%%%%%%%%%%%%%%%%%%%%%%%%%%%%%%%%%%%%%%%%%%%%%%%%%%%%%%%%%%%%%%%%%%%%%%%%%%%%%%%%
\subsection{Assurance Tests Based on Multiple Objectives}\label{subsec:multiple.objectives}
%%%%%%%%%%%%%%%%%%%%%%%%%%%%%%%%%%%%%%%%%%%%%%%%%%%%%%%%%%%%%%%%%%%%%%%%%%%%%%%%%%%%%%%%%%%%%%%%%%%%%%%%%%%%%%%%%%%%%

Note that for simplicity of discussion, in the following sections, we employ abbreviated notation including CR to represent the PCR, and PR to signify the PPR. After selecting the risk criteria, the development of assurance test plans depends on the degree of risk that practitioners are willing to accept based on their specific applications and available resources. For example, for zero failure tests, a test is deemed to be successful if no failure is observed $(y=0)$ during the test period $(\tauh,\tauh+\taut]$. The zero-failure test plans have been popular, as they require minimal number of test units $\nt$ while controlling the CR. However, these tests are often associated with high PR, and low AP. Hence, it is necessary to develop the assurance test plans based on multiple objectives. Moreover, it is also important to understand the trade-offs between each objective and then make a balanced decision in accordance with the specific goals within a set of vehicle test plans.

In general, suppose we have multiple objective functions and $\zvec$ is our decision vector. A solution $\zvec_{1}$ is said to Pareto dominate another solution $\zvec_{2}$ if (i) solution $\zvec_{1}$ is as good as $\zvec_{2}$ based on all objectives, and (ii) solution $\zvec_{1}$ is strictly better than $\zvec_{2}$ based on at least one objective. The non-dominated solution set consists of all the solutions that are not dominated by any other members. The Pareto front approach searches for all the non-dominated solutions based on considering multiple objectives. The Pareto front consists of all the non-dominated points mapped from the Pareto optimal solutions into the criterion space.

When multiple objectives are of interest in test planning, due to the trade-offs between the criteria, there is often no universal solution to simultaneously optimize all criteria under consideration. Under this situation, to select the best test plan in a specific scenario, we need to prioritize the competing objectives and make a tailored decision to best match the goals. To obtain a sensible test plan, we want to control the CR or the PR to be:
\begin{align}\label{eqn:Pareto.front.control.CR.PR}
\Pr(m(s,t) \geq m_{1} \vert \text{Test is Passed} ) & \leq  \alpha_c \\\nonumber
\Pr(m(s,t) \leq m_{0} \vert \text{Test is Failed} ) & \leq  \alpha_p,
\end{align}
where $\alpha_c$ and $\alpha_p$ represent the user-defined thresholds for the consumer's and the producer's risks, respectively.

This paper adapts the Pareto front approach proposed by \shortciteN{lu2016multiple}, to identify a collection of non-dominate test plans considering multiple risk criteria. Considering CR is often of the most importance among all the risks, we identify the Pareto front among solutions with acceptable CR values, i.e., we seek to:
\begin{align} \label{eqn:Pareto.front.constraint.CR}
&\text{minimize }\Pr(m(s,t) \leq m_{0} \vert \text{Test is Failed} )\\\nonumber
&\text{maximize }\Pr(\text{Test is Passed} )\\\nonumber
&\text{minimize }  \tau \text{ or }\taut \\\nonumber
\text{s.t. } &\Pr(m(s,t) \geq m_{1} \vert \text{Test is Passed} )\leq  \alpha_c. \end{align}

The Pareto front approach can help eliminate non-competitive options from the decision-making, ultimately facilitating more informed decisions. We consider four criteria for all potential test plans under the HPP and NHPP models, respectively. These criteria include (1) CR, (2) PR, (3) the total vehicle days $\tau$ for the HPP model or the testing period per vehicle $\taut$ for the NHPP model, and (4) AP. Next, we will illustrate the detailed decision-making process to select the most suitable demonstration test plan, taking into account multiple criteria at the same time. For each model, we will explore the interrelationships among the multiple risk criteria. Then, we will identify a set of non-dominating test plans by applying the Pareto front method. Finally, we will make further recommendations on how to select the best test plan for execution from the Pareto front to match different user priorities.

%%%%%%%%%%%%%%%%%%%%%%%%%%%%%%%%%%%%%%%%%%%%%%%%%%%%%%%%%%%%%%%%%%%%%%%%%%%%%%%%%%%%%%%%%%%%%%%%%%%%%%%%%%%%%%%%%%%%%
\section{Test Plans Based on Homogeneous Poisson Process}\label{sec:HPP.main}

%%%%%%%%%%%%%%%%%%%%%%%%%%%%%%%%%%%%%%%%%%%%%%%%%%%%%%%%%%%%%%%%%%%%%%%%%%%%%%%%%%%%%%%%%%%%%%%%%%%%%%%%%%%%%%%%%%%%%
\subsection{Risk Criteria}\label{subsec:HPP.risk.criteria}
%%%%%%%%%%%%%%%%%%%%%%%%%%%%%%%%%%%%%%%%%%%%%%%%%%%%%%%%%%%%%%%%%%%%%%%%%%%%%%%%%%%%%%%%%%%%%%%%%%%%%%%%%%%%%%%%%%%%%

In this paper, we consider the AV test planning after the reliability growth process. First, we consider the case when the failure intensity of the system can no longer be reduced and remains at a constant rate. We use the average intensity $m(s,t)$ as the reliability metric for the test planning. Following \eqref{eqn:reliability_metric}, for the HPP model, $m(s,t)=m$, which remains constant throughout the testing period. We evaluate the performance of any $(\nt, \taut, c)$ test plan based on the (1) CR, (2) PR, (3) $\tau$, and (4) AP, where $\tau=\nt\taut$ representing the total test vehicle days. We demonstrate a comprehensive evaluation of all possible test plans under the HPP model, and examine the inter-relationship between the test criteria.

%%%%%%%%%%%%%%%%%%%%%%%%%%%%%%%%%%%%%%%%%%%%%%%%%%%%%%%%%%%%%%%%%%%%%%%%%%%%%%%%%%%%%%%%%%%%%%%%%%%%%%%%%%%%%%%%%%%%%
\subsection{Planning Values}\label{subsec:HPP.planning.values}
%%%%%%%%%%%%%%%%%%%%%%%%%%%%%%%%%%%%%%%%%%%%%%%%%%%%%%%%%%%%%%%%%%%%%%%%%%%%%%%%%%%%%%%%%%%%%%%%%%%%%%%%%%%%%%%%%%%%%

As discussed in Section~\ref{subsec:stat.models}, determining the BIF is the initial step in calculating the event intensity function for unit $i$ at time $t$ under the HPP. In this paper, we consider the Weibull reliability growth model, where the BIF can be expressed as:
\begin{align} \label{eqn:Weibull.BIF}
&\lambda_{0}(t;\thetavec)=\theta_{1}\theta_{2}\theta_{3}t^{(\theta_{3}-1)}\exp(-\theta_{2}t^{\theta_{3}}),
\end{align}
and the CBIF takes the form:
\begin{align} \label{eqn:Weibull.CBIF}
\Lambda_{0}(t;\thetavec)=\theta_{1}[1-\exp(-\theta_{2}t^{\theta_{3}})].
\end{align}
In the above expressions, $\thetavec=(\theta_{1},\theta_{2},\theta_{3})'$ and $\theta_{1}>0, \theta_{2}>0, \theta_{3}>0$. To evaluate the Bayesian risk criteria, we use the posterior distribution of $\thetavec$ conditioned on the historical CA DMV test data to compare our knowledge about the planning parameters prior to the test planning.

To fully explore the relationships between the four criteria with the test plan parameters $(\nt,\taut,c)$, we explore $\taut$ ranging from $20$ to $365$ days for a total of possible $10$ test units, which results in $\tau$ values ranging between $200$ to $3650$ vehicle days. We chose the average daily driving distance to be $0.21$ k-miles, for the testing and demonstration periods. Also, we examine cases where the number of maximum allowable failures $c$ ranges between $0$ and $50$. And we set the reliability requirement at $m_{1}=0.016$ and $m_{0}=0.013$ (i.e., the maximum acceptable failure intensity for the consumer is set at $0.016$ and the minimum rejectable failure intensity for the producer is defined at $0.013$). The values of the planning parameters are drawn from the posterior distribution obtained based on the CA DMV test data from 2017 to 2019.

%%%%%%%%%%%%%%%%%%%%%%%%%%%%%%%%%%%%%%%%%%%%%%%%%%%%%%%%%%%%%%%%%%%%%%%%%%%%%%%%%%%%%%%%%%%%%%%%%%%%%%%%%%%%%%%%%%%%%
\subsection{Examples}\label{subsec:HPP.example}
%%%%%%%%%%%%%%%%%%%%%%%%%%%%%%%%%%%%%%%%%%%%%%%%%%%%%%%%%%%%%%%%%%%%%%%%%%%%%%%%%%%%%%%%%%%%%%%%%%%%%%%%%%%%%%%%%%%%%
For each test plan $(\nt,\taut,c)$ under the HPP framework, all criteria values are calculated based on \eqref{eqn:pcr.hpp} to \eqref{eqn:ap.hpp}. Before choosing the final test plan, it is helpful for us to investigate the relationships between different risk criteria among all the test plans. This exploration provides an improved understanding of the trade-offs between the test criteria and how they are interconnected.

\begin{figure}[ht!]
\begin{center}
\begin{tabular}{cc}
\includegraphics[width=0.50\textwidth]{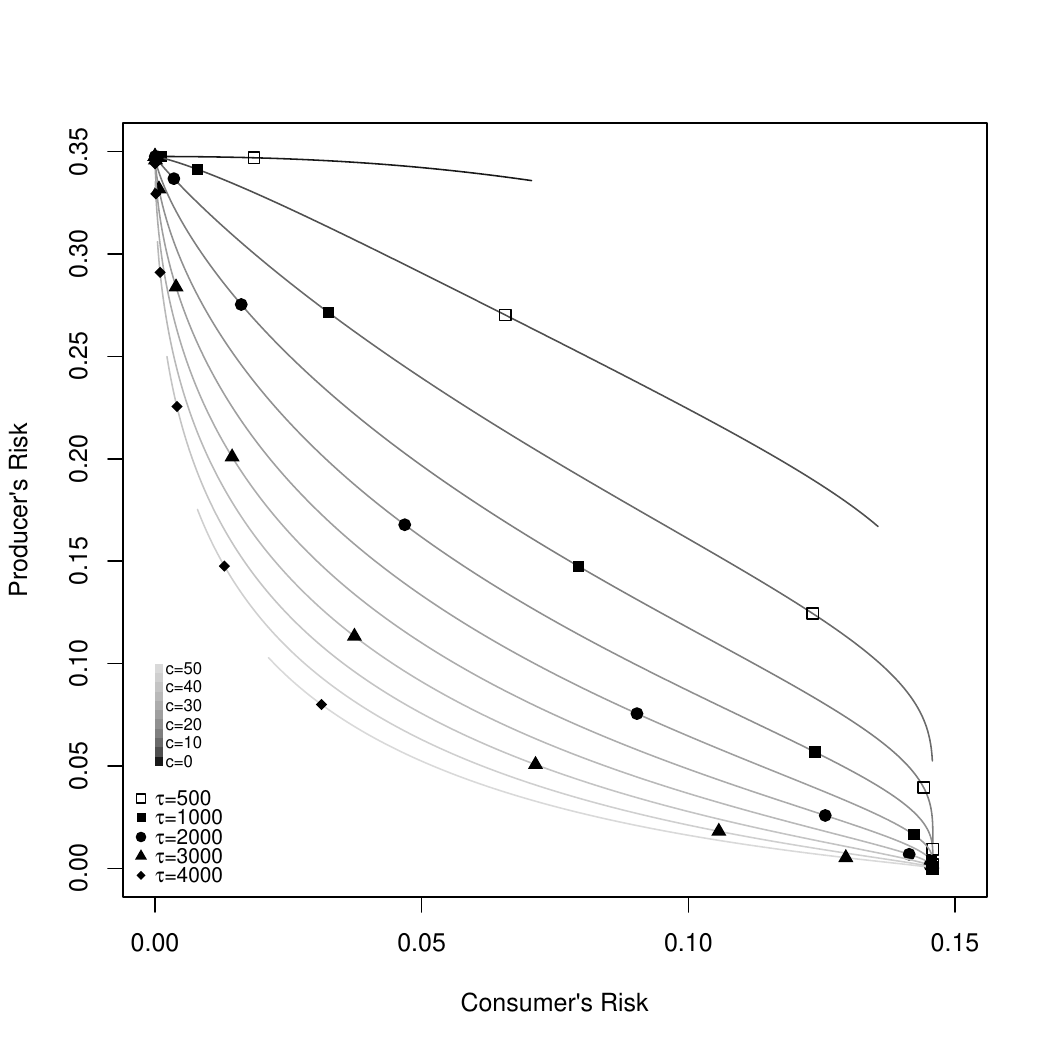} &
\includegraphics[width=0.50\textwidth]{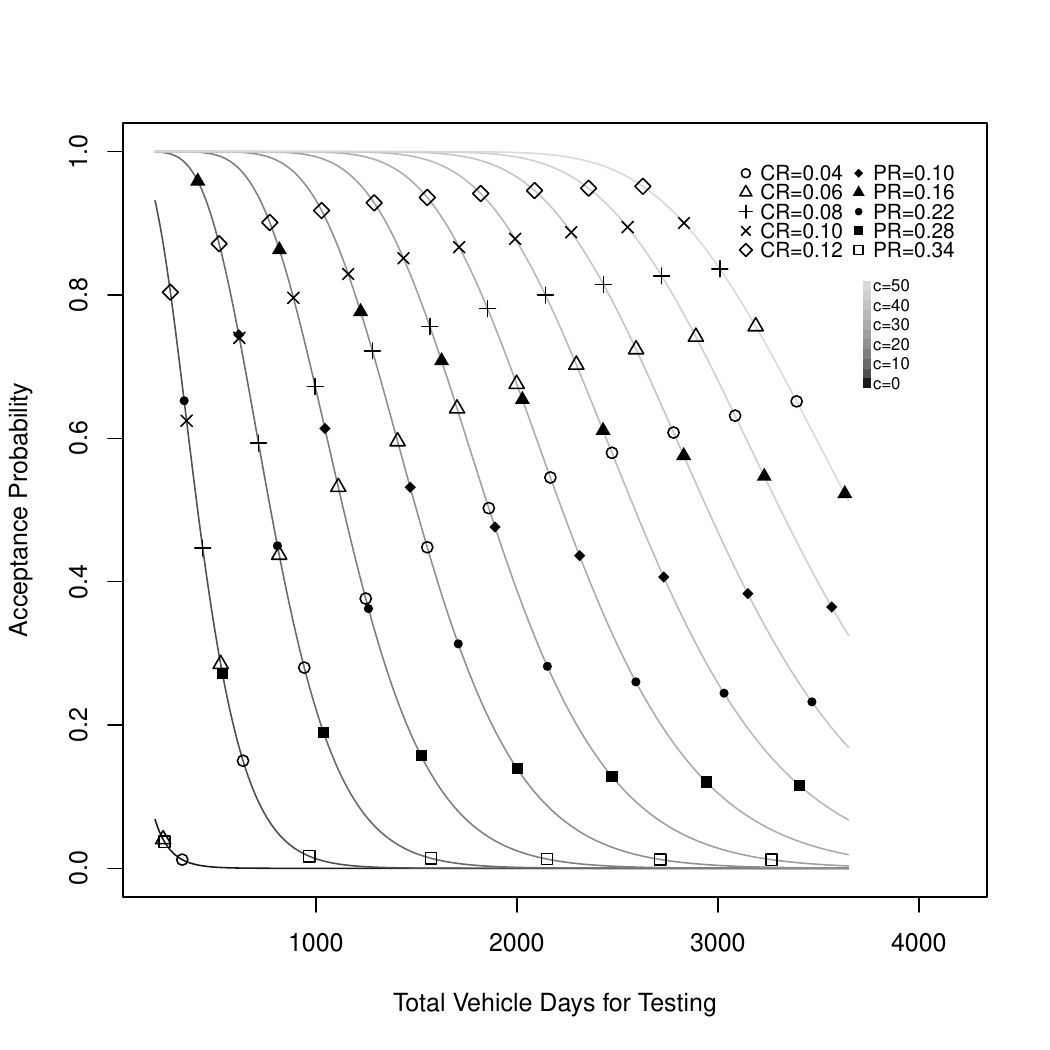}\\
(a) CR vs. PR    & (b) $\tau$ vs. AP      \\
\includegraphics[width=0.50\textwidth]{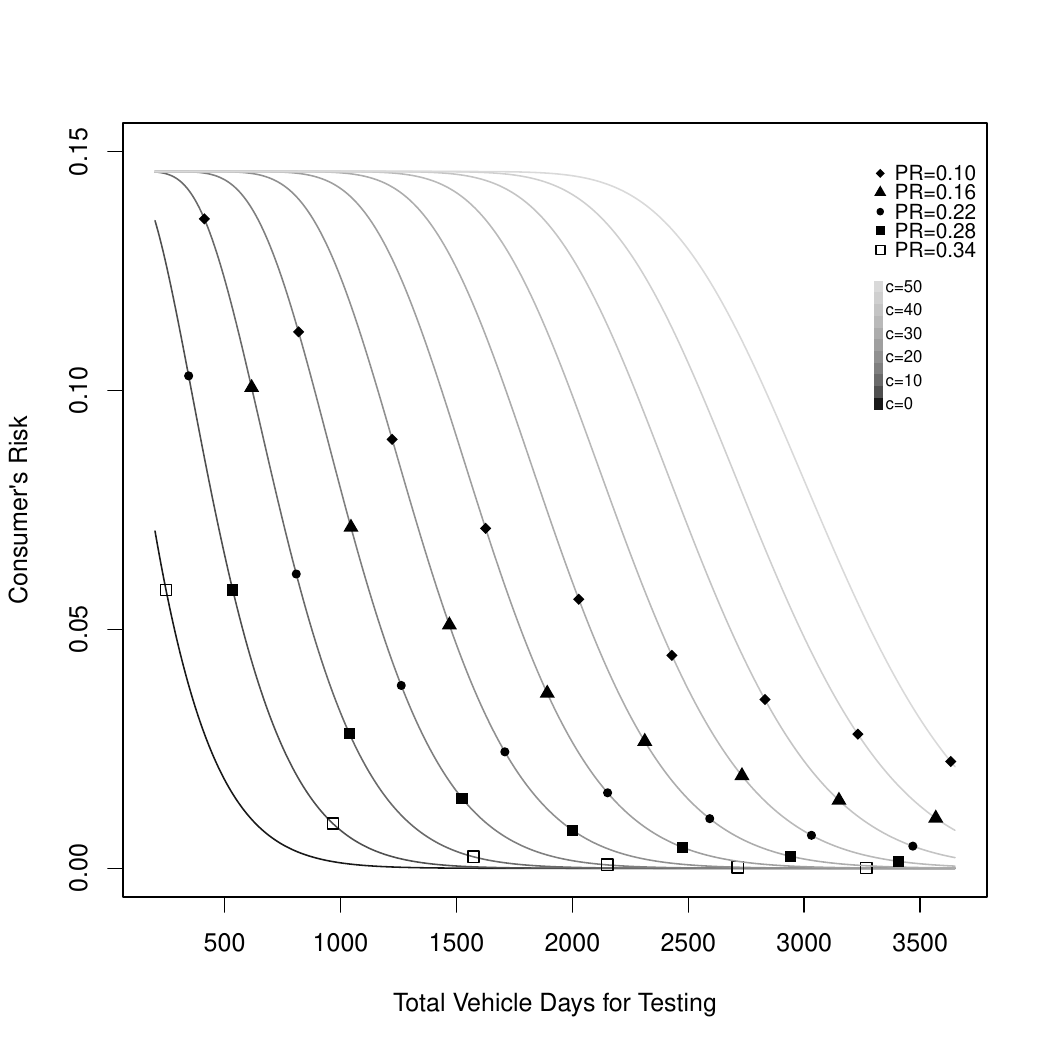} &
\includegraphics[width=0.50\textwidth]{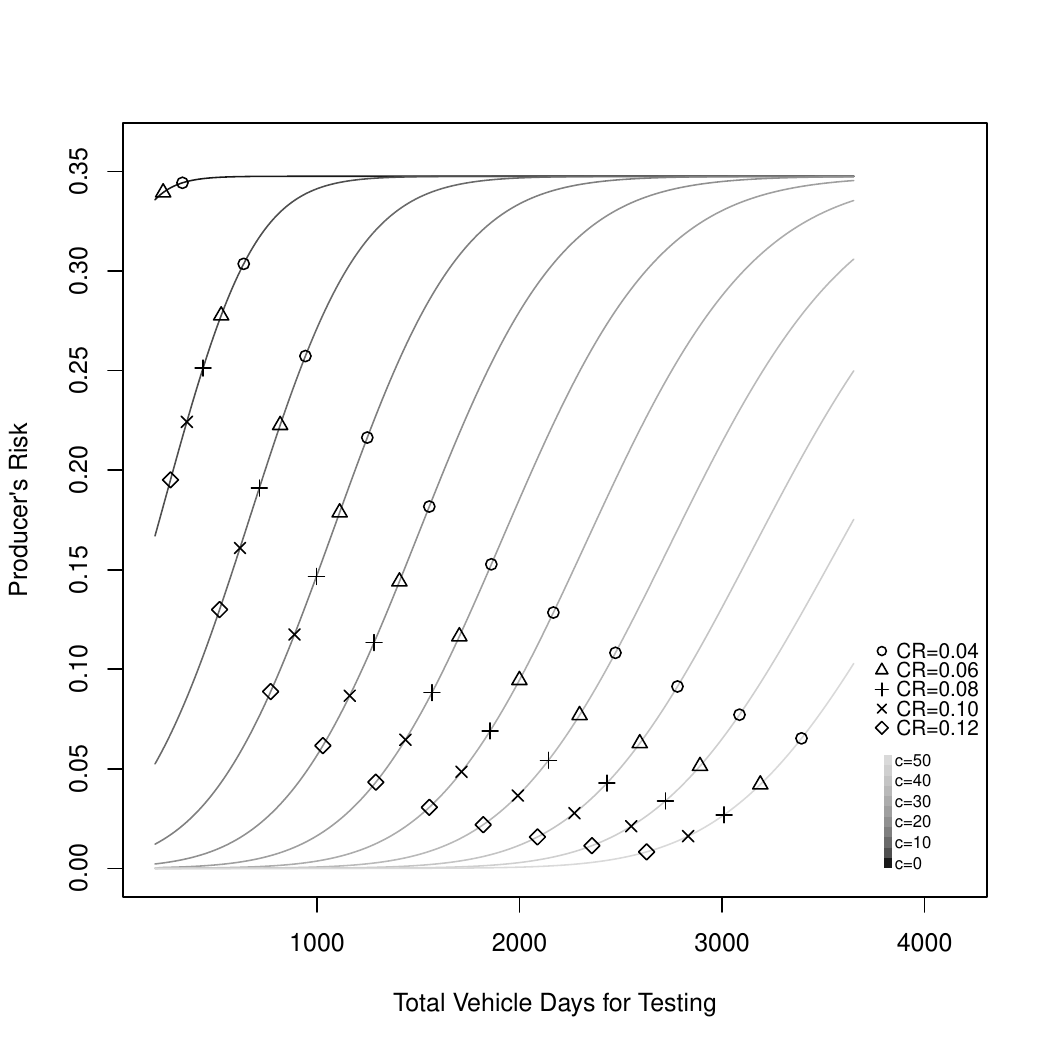}\\
(c) $\tau$ vs. CR   & (d) $\tau$ vs. PR
\end{tabular}
 \caption{For the representative sample of test plans explored under the HPP model, the plots show the inter-relationships between the CR, PR, AP, and $\tau$. Within each panel, test plans with the same value of $c$ align on the same curve. The dark to light shades indicate small to large $c$ values in the range of $[0,50]$. Different symbols represent selected representative values for the other criteria.} \label{fig:all.possible.tests.hpp}
\end{center}
\end{figure}

Figure~\ref{fig:all.possible.tests.hpp} shows the performance of the representative samples from the test plans within the explored range based on the four criteria discussed in Section~\ref{subsec:HPP.risk.criteria} under the HPP model. Figure~\ref{fig:all.possible.tests.hpp}(a) shows a plot of  CR and PR for the sampled test plans. We use curves in different grey shades representing different $c$ values, where the darker colors correspond to smaller values. Different symbols are used to represent different test durations. We can observe some obvious patterns. First, there exists a strong trade-off between CR and PR at any fixed value of $c$. Particularly, as CR increases, PR decreases at a declining rate with the same $c$ value. This indicates the PR can be improved at the cost of increasing CR. However, there is a diminishing return as the improvement in PR becomes smaller when CR gets larger. Second, we can see that as we increase $c$, both PR and CR can be simultaneously reduced by increasing $\tau$. This is revealed from observing lighter gray curves towards to the bottom left corner, with reduced CR and PR. Note that the improvements on the CR and PR values also reduce as $c$ increases. Third, we notice that the range of CR across the explored test plans is between $0$ and $0.15$. Meanwhile, the PR has a slightly broader range than CR, from $0$ to $0.35$, indicating more test plans with potentially higher PR than CR. In addition to the curves shown in Figure~\ref{fig:all.possible.tests.hpp}(a), we also explored some specific test plans with $\tau=500, 1000, 1500, 2500,$ and $3000$ to explore the effects of changing $\tau$. We can see that at any fixed $\tau$, we can reduce PR by increasing the maximum allowable failures. Also, given any fixed $c$, increasing $\tau$ will reduce CR but increase PR.

Figure~\ref{fig:all.possible.tests.hpp}(b) shows the relationship between AP and $\tau$ for the representative test plans at different levels of $c$. Note that here we use $\tau$ as one of the criteria under the HPP model, considering that $\nt$ potentially can change. At each fixed $c$ value, the AP decreases as the total test vehicle days $\tau$ increases. This indicates that given a fixed maximum number of failures $(c)$, the longer the total test vehicle days (either testing more vehicles or for a longer test duration), the smaller chance there is to pass the test. On the other hand, given a fixed total test vehicle days $(\tau)$, the chance of accepting the test increases as a larger $c$ is allowed. In addition, different symbols represent the test plans at controlled the CR levels $(0.04,0.06,0.08,0.10$, and $0.12)$ or controlled the PR levels $(0.10,0.16,0.22,0.28$, and $0.34)$. We can see that, at a  fixed $\tau$, test plans with higher AP are generally associated with larger CR and smaller PR.

Figure~\ref{fig:all.possible.tests.hpp}(c) shows the relationship between CR and $\tau$. Additionally, we have highlighted selected PR values at $0.10, 0.16, 0.22, 0.28$, and $0.34$. Similar to Figure~\ref{fig:all.possible.tests.hpp}(b), at a fixed value of $c$, CR decreases and PR increases as the total test vehicle days increase. While at a fixed $\tau$, we can reduce CR at the cost of increasing PR by increasing $c$. For the sampled test plans at controlled PR values, we can reduce CR by increasing $\tau$ and $c$.

Figure~\ref{fig:all.possible.tests.hpp}(d) shows the relationship between PR and $\tau$, and highlights the test plans with controlled CR values at $0.04,0.06,0.08,0.10$, and $0.12$. We can see that at a fixed $c$ value, the PR can be reduced while raising CR by reducing the total test vehicle days $\tau$. On the other hand, at a fixed $\tau$, the PR can be improved at the cost of CR if we allow more failures to pass the test. When controlling the CR, the PR can be reduced by allowing larger $\tau$ and $c$ values.

To summarize, CR and PR have the most severe trade-off among all the evaluated test criteria. When one of the $c$ or $\tau$ is fixed, we can adjust the other parameter to reduce the one of risk criteria, while sacrificing the performance of the other. To reduce both CR and PR, we need to increase $c$ and $\tau$ at the same time. However, this will increase the total cost ($\tau$) and decrease the chance of passing the test.

To select a test plan, we consider CR as the most important among the four criteria, and we aim to control CR at or below $0.086$. We consider all the test plans that meet this primary objective. Then we remove inferior solutions by finding the Pareto front with the set of non-dominated solutions based on the remaining three criteria (PR, AP, and $\tau$).

\begin{figure}%[ht!]
\begin{center}
\includegraphics[width=0.65\textwidth]{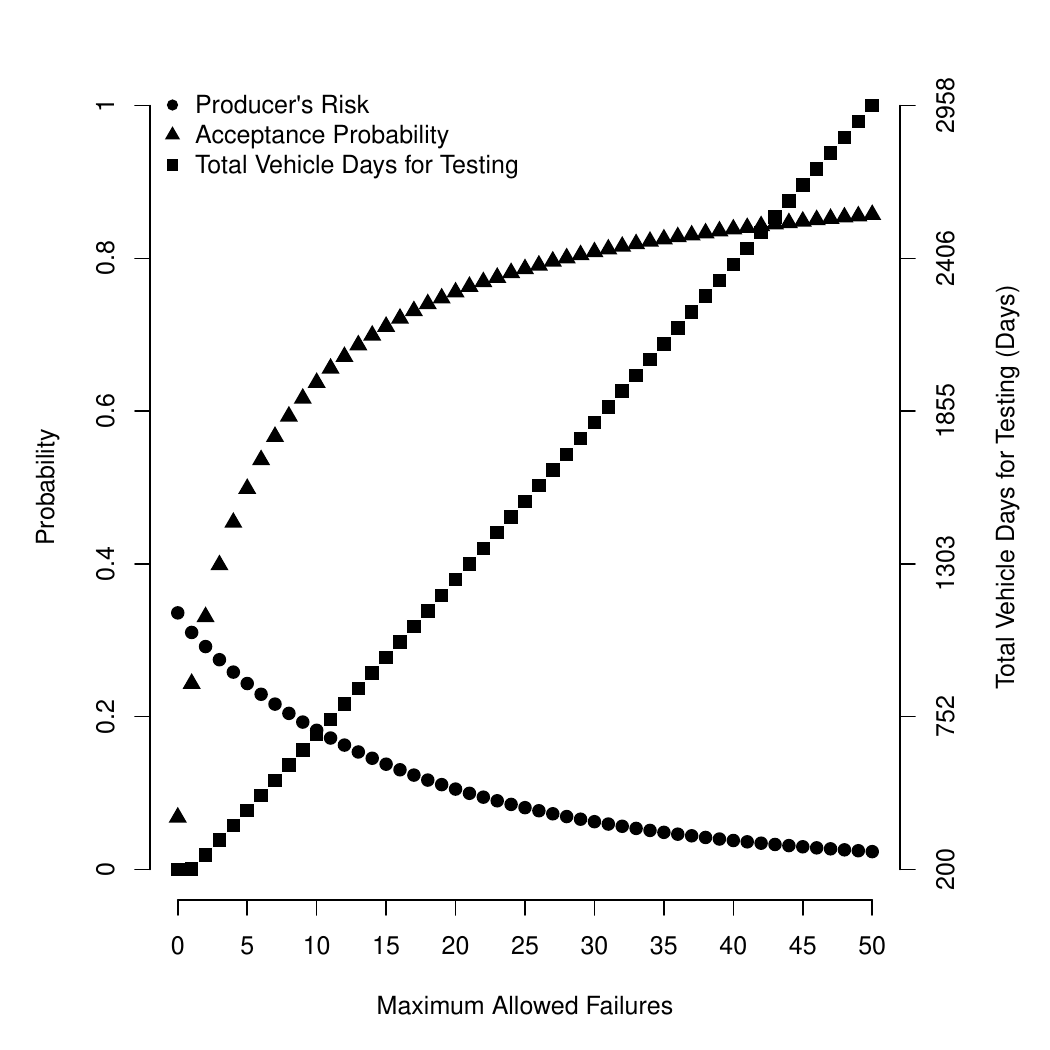}
\caption{The trade-off plot for Pareto front of the optimal test plans under HPP given CR being controlled at or below 0.086. Note that there are 51 choices on the three criteria Pareto front based on the PR, AP, and $\tau$, considering the constraint that CR does not exceed 0.086. The left axis represents PR and AP, while the scale on the right indicates the total testing time. Each symbol denotes the respective Pareto optimal solutions for the remaining three criteria, for varying $c$ values within the $[0,50]$ range, arranged from left to right in ascending order.}
\label{fig:PF.hpp}
\end{center}
\end{figure}

Figure~\ref{fig:PF.hpp} shows the performance of all the test plans on the Pareto front based on PR, AP and $\tau$, subject to $\text{CR} \leq 0.086$. From Figure~\ref{fig:PF.hpp}, it offers a direct method to simplify the test plan selection, considering the constraint of CR. The Pareto front solutions ultimately consist of $51$ test plans corresponding to different $c$ values. This suggests that for a given $c$, there is a universal optimal test plan when optimizing PR, AP, and $\tau$ simultaneously. 

To better understand Figure~\ref{fig:PF.hpp}, we can see that the Pareto front with the set of non-dominated solutions is organized left to right with an increasing $c$ value. The left vertical axis scales from $0$ to $1$, serving as a measure for PR and AP. The right vertical axis ranges from $200$ to $2833$ and is used for measuring cost based on the total test vehicle days $\tau$. Regarding the trade-off among all the other three criteria, based on the competing solutions, we can find that the total testing vehicle days increases from $20$ days at $c=0$ to $2833$ days at $c=50$, while PR reducing from approximately $0.31$ to just below $0.02$, and AP increasing from roughly $0.25$ to above $0.90$. This indicates that by increasing both $\tau$ and $c$, we can substantially improve PR and AP.

By using this trade-off plot which includes the Pareto front with the set of non-dominated solutions, users can make straightforward and informed decisions. These decisions can be made based on factors including the available budget, affordable total test vehicle time, risk tolerance level, or the minimum acceptance probability of passing a test plan. For example, if PR is considered more important among the remaining criteria and the producer cannot accept a test plan with PR higher than $0.1$, then the best plan is to test $1302$ vehicle days in total for a possible $10$ test units and allow for up to $21$ failures. This will result in a test plan with (1) CR at $0.086$, (2) PR at $0.099$, and (3) AP at $0.763$. In contrast, if the budget allows only up to $1000$ vehicle test days, the optimal plan is to test $965$ total vehicle days with a maximum of $15$ failures. This test plan results in testing (1) CR at $0.086$, (2) PR at $0.138$, and (3) AP at $0.710$. Alternatively, we might have a more strict limitation for the maximum allowable failure. For example, if we can allow no more than $10$ failures, then the best plan is to test for $687$ total vehicle days, with up to $10$ failures. This will result in (1) CR at $0.086$, (2) PR at $0.182$, and (3) AP at $0.637$. Note this is only to illustrate the decision-making process. The procedure can be flexible to adapt to different user priorities. The selected test plan would also vary with different user priorities, the choices of the prior distributions, and the reliability requirements on the average failure intensity.

%%%%%%%%%%%%%%%%%%%%%%%%%%%%%%%%%%%%%%%%%%%%%%%%%%%%%%%%%%%%%%%%%%%%%%%%%%%%%%%%%%%%%%%%%%%%%%%%%%%%%%%%%%%%%%%%%%%%%
\section{Test Plans Based on Non-homogeneous Poisson Process}\label{sec:NHPP.main}
%%%%%%%%%%%%%%%%%%%%%%%%%%%%%%%%%%%%%%%%%%%%%%%%%%%%%%%%%%%%%%%%%%%%%%%%%%%%%%%%%%%%%%%%%%%%%%%%%%%%%%%%%%%%%%%%%%%%%
\subsection{Risk Criteria}\label{subsec:NHPP.risk}
%%%%%%%%%%%%%%%%%%%%%%%%%%%%%%%%%%%%%%%%%%%%%%%%%%%%%%%%%%%%%%%%%%%%%%%%%%%%%%%%%%%%%%%%%%%%%%%%%%%%%%%%%%%%%%%%%%%%%
Next, we consider the case where the failure intensity of the system varies throughout the testing period.  We use the average intensity as the reliability metric for the test planning, as discussed in Section~\ref{subsec:reliability.metrics}. The calculation of $m(s,t)$ for the NHPP model is based on \eqref{eqn:reliability_metric}. Under the NHPP model, suppose the goal is to demonstrate the reliability performance at the end of the demonstration period ($\tauh + \taud$). All the test parameters including (1) $\tauh + \taud$, (2) $\tauh + \taut$, (3) $m_0$ and (4) $m_1$ are specified based on the test objectives. We focus on the four aspects of the test plan performance including (1) CR, (2) PR, (3) AP, and (4) $\taut$.

%%%%%%%%%%%%%%%%%%%%%%%%%%%%%%%%%%%%%%%%%%%%%%%%%%%%%%%%%%%%%%%%%%%%%%%%%%%%%%%%%%%%%%%%%%%%%%%%%%%%%%%%%%%%%%%%%%%%%
\subsection{Planning Values}\label{subsec:NHPP.plannning.values}
%%%%%%%%%%%%%%%%%%%%%%%%%%%%%%%%%%%%%%%%%%%%%%%%%%%%%%%%%%%%%%%%%%%%%%%%%%%%%%%%%%%%%%%%%%%%%%%%%%%%%%%%%%%%%%%%%%%%%
Under the NHPP model, which has non-constant intensity function over the testing period, we examine two different scenarios: one involves a single test vehicle and the other involves multiple test vehicles. Before discussing the specific parameter settings for these two scenarios, we define the form of the CBIF under the NHPP model based on \eqref{eqn:Weibull.CBIF}.

To fully investigate the inter-relationships among the four test criteria, first we consider a single test unit scenario for a test duration of one year (e.g., $\taut=365$ days). The demonstration period is set for two years, with $\taud=730$ days. The average daily mileage is set at $0.20$ k-miles for both the testing and demonstration periods. For testing a single vehicle with  $\nt=1$, we explore the range of $c$ between 0 to 5. The reliability requirement is set at $m_1=0.0125$ and $m_0=0.009$ (i.e., the maximum acceptable failure intensity for the consumer is set at 0.0125, and the minimum rejectable failure intensity for the producer is established at 0.009). In addition, the posterior distribution of the model parameters $\pi(m\vert \DATA)$, derived from the CA DMV dataset from 2017 to 2019, will be used as the priors in calculating the risk criteria.

For the fleet testing scenario, we use the same settings for (1) $\taut$, (2) $\taud$, (3) average daily mileage for both testing and demonstration periods, (4) $\pi(m)$ and (5) $m_0$. However, we adjust the following settings. First, we choose to explore the scenario with $n_{\text{t}}=5$, and
$m_1=0.0132$. Also, the number of allowable failures is set to range from 0 to 25, for testing 5 vehicles simultaneously.

%%%%%%%%%%%%%%%%%%%%%%%%%%%%%%%%%%%%%%%%%%%%%%%%%%%%%%%%%%%%%%%%%%%%%%%%%%%%%%%%%%%%%%%%%%%%%%%%%%%%%%%%%%%%%%%%%%%%%
\subsection{Examples}\label{subsec:NHPP.examples}
%%%%%%%%%%%%%%%%%%%%%%%%%%%%%%%%%%%%%%%%%%%%%%%%%%%%%%%%%%%%%%%%%%%%%%%%%%%%%%%%%%%%%%%%%%%%%%%%%%%%%%%%%%%%%%%%%%%%%

\begin{figure}[ht!]
\begin{center}
\begin{tabular}{cc}
\includegraphics[width=0.50\textwidth]{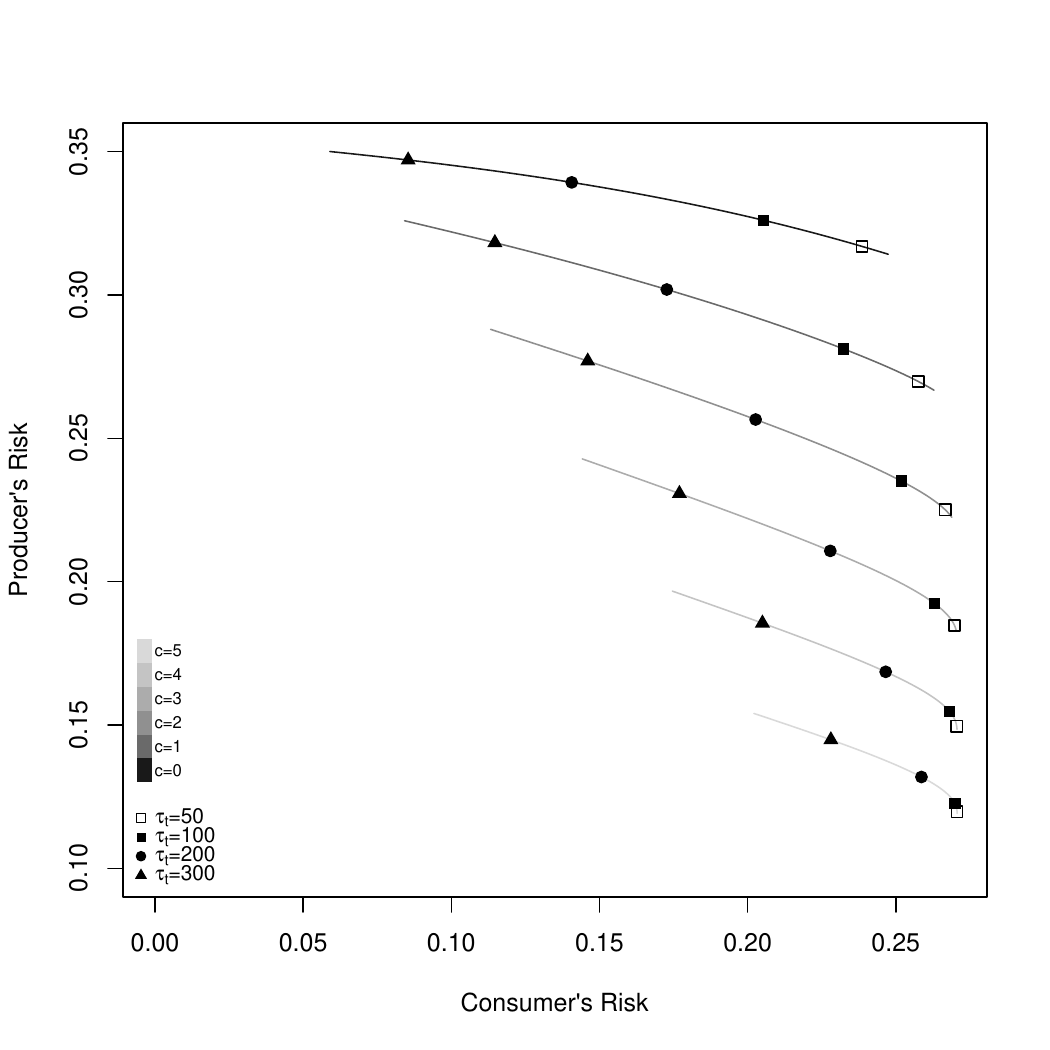} &
\includegraphics[width=0.50\textwidth]{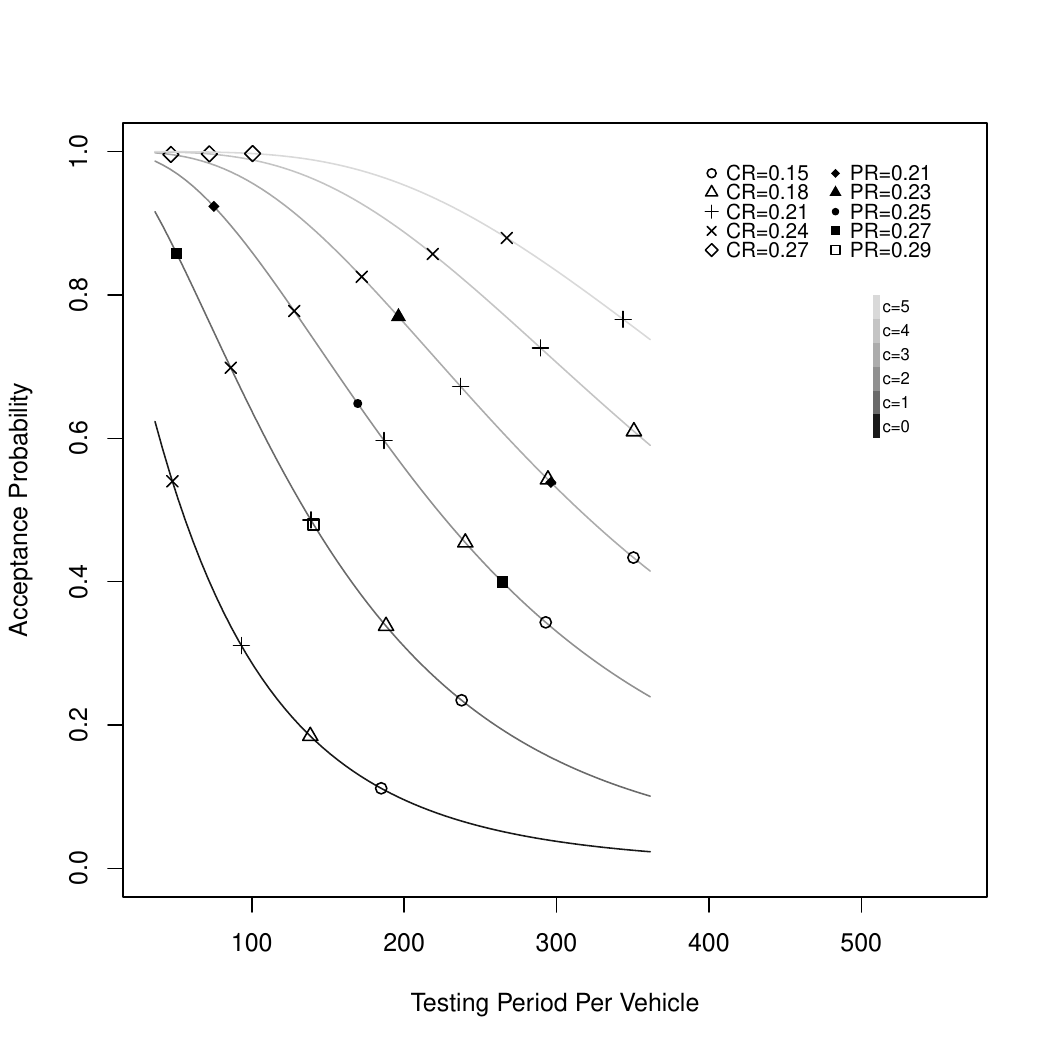}\\
(a) CR vs. PR    & (b) $\taut$ vs. AP      \\
\includegraphics[width=0.50\textwidth]{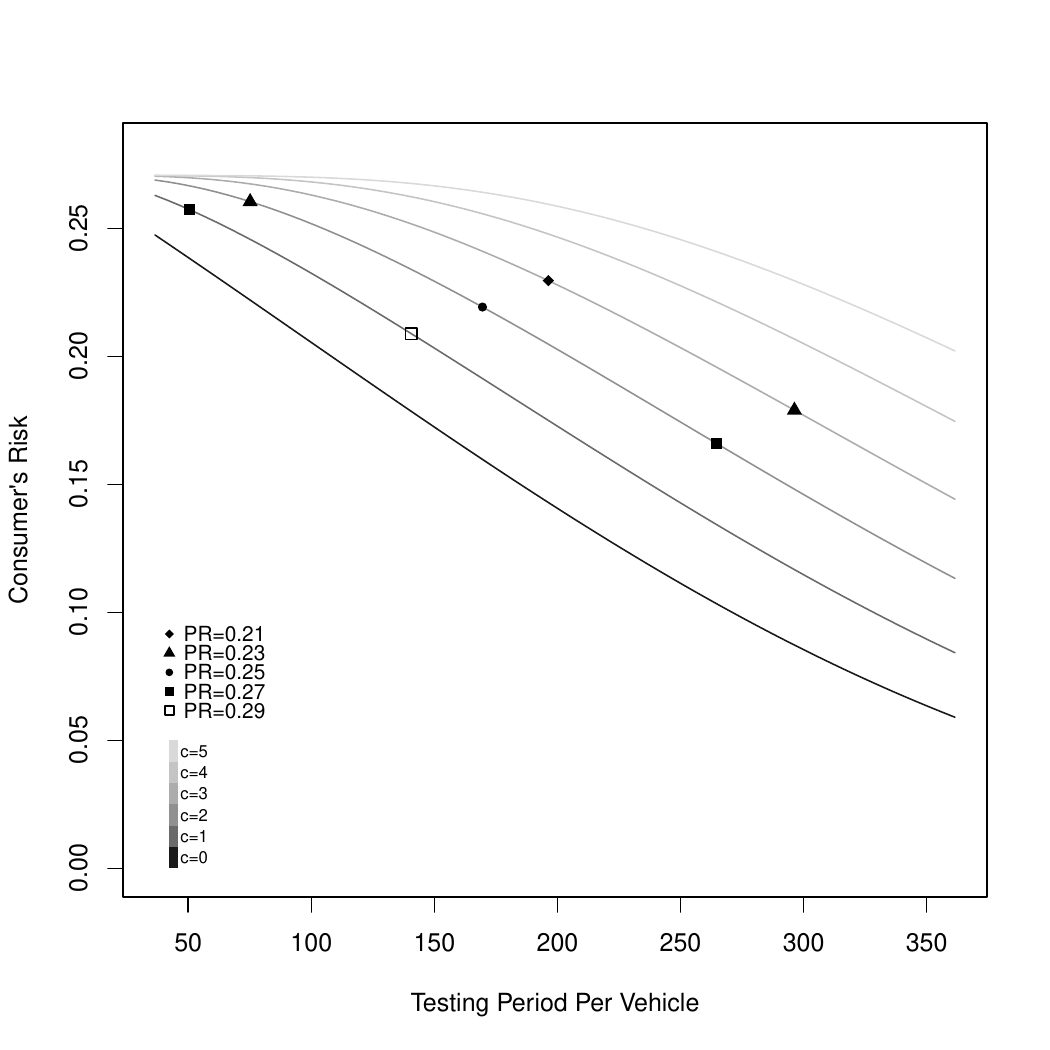} &
\includegraphics[width=0.50\textwidth]{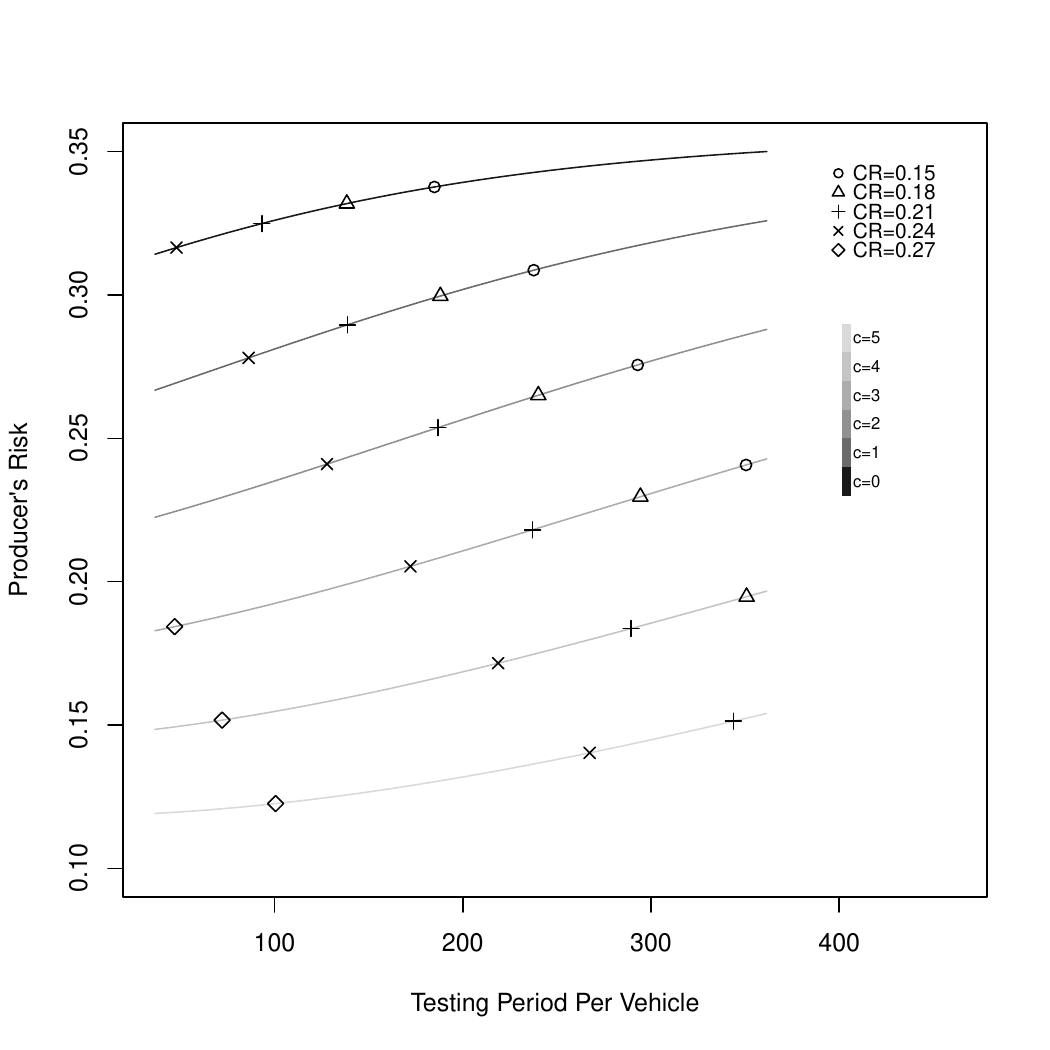}\\
(c) $\taut$ vs. CR   & (d) $\taut$ vs. PR
\end{tabular}
\caption{All possible test plans for the single vehicle test under the NHPP. The above plots shows the inter-relationships between CR, PR, AP, and $\taut$. In each plot, test plans with identical $c$ values are on the same curve distinguished by gradient gray shades. These shades transition from darker to lighter to represent increasing $c$ values within the $[0,5]$ range. Similarly, each symbols indicates selected representative symbols for other criteria. }
\label{fig:all.possible.tests.nhpp.single}
\end{center}
\end{figure}

\begin{figure}[ht!]
\begin{center}
\begin{tabular}{cc}
\includegraphics[width=0.50\textwidth]{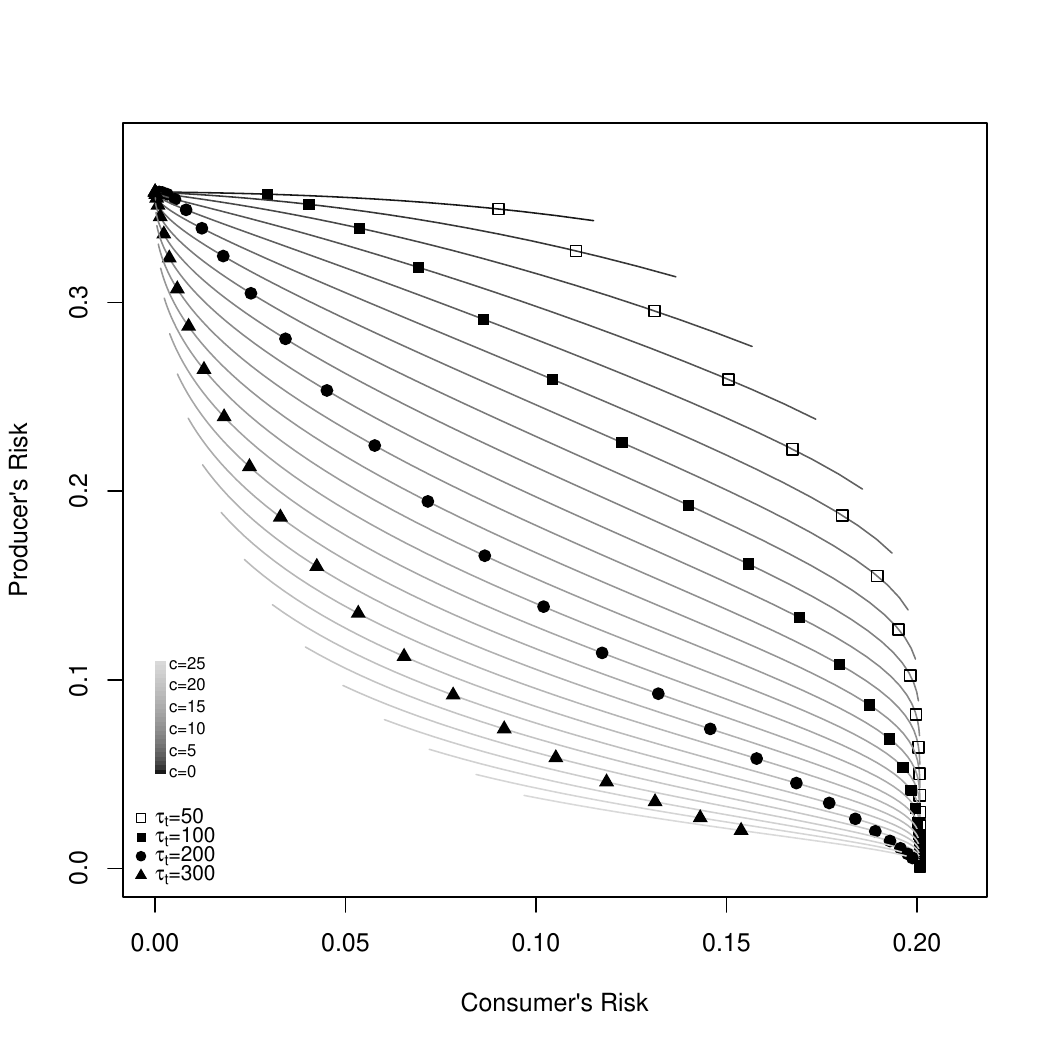} &
\includegraphics[width=0.50\textwidth]{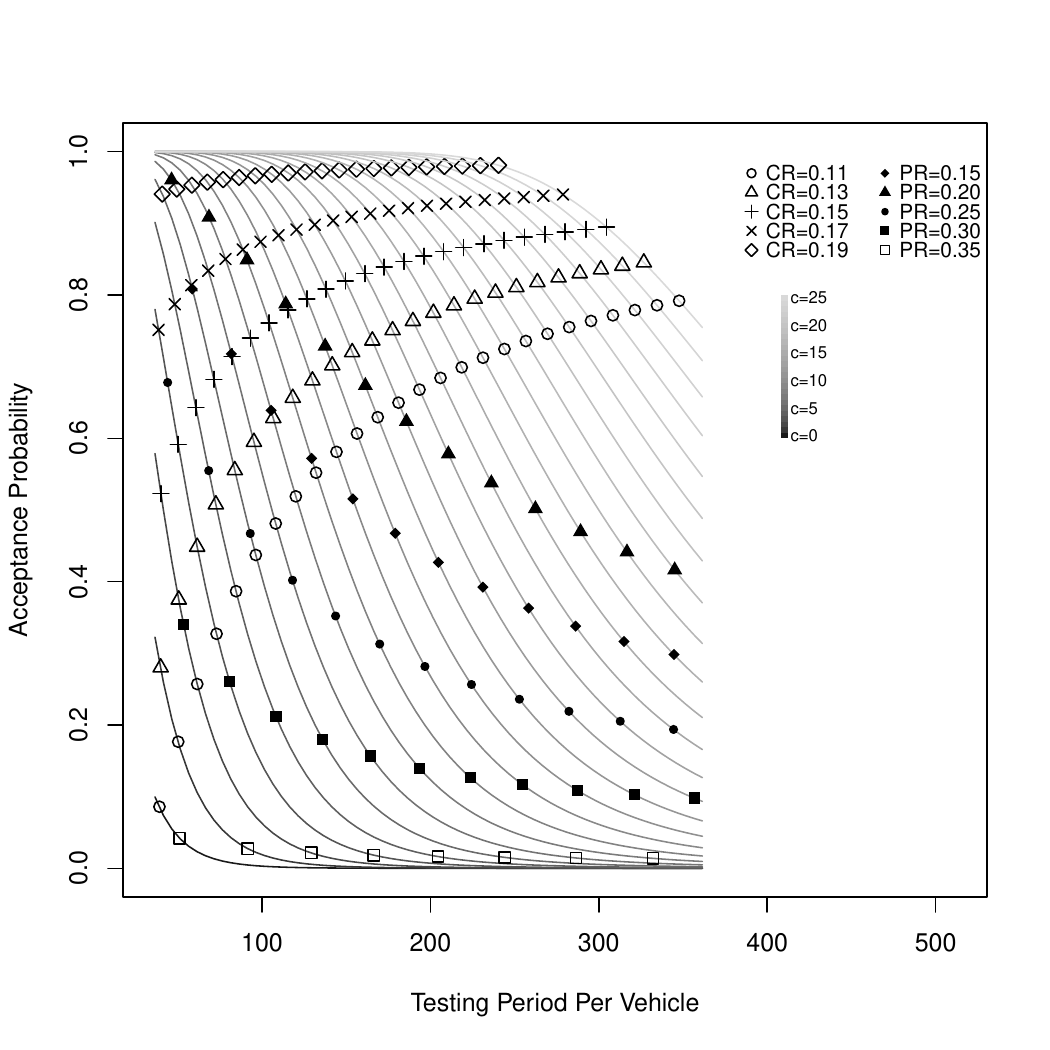}\\
(a) CR vs. PR    & (b) $\taut$ vs. AP     \\
\includegraphics[width=0.50\textwidth]{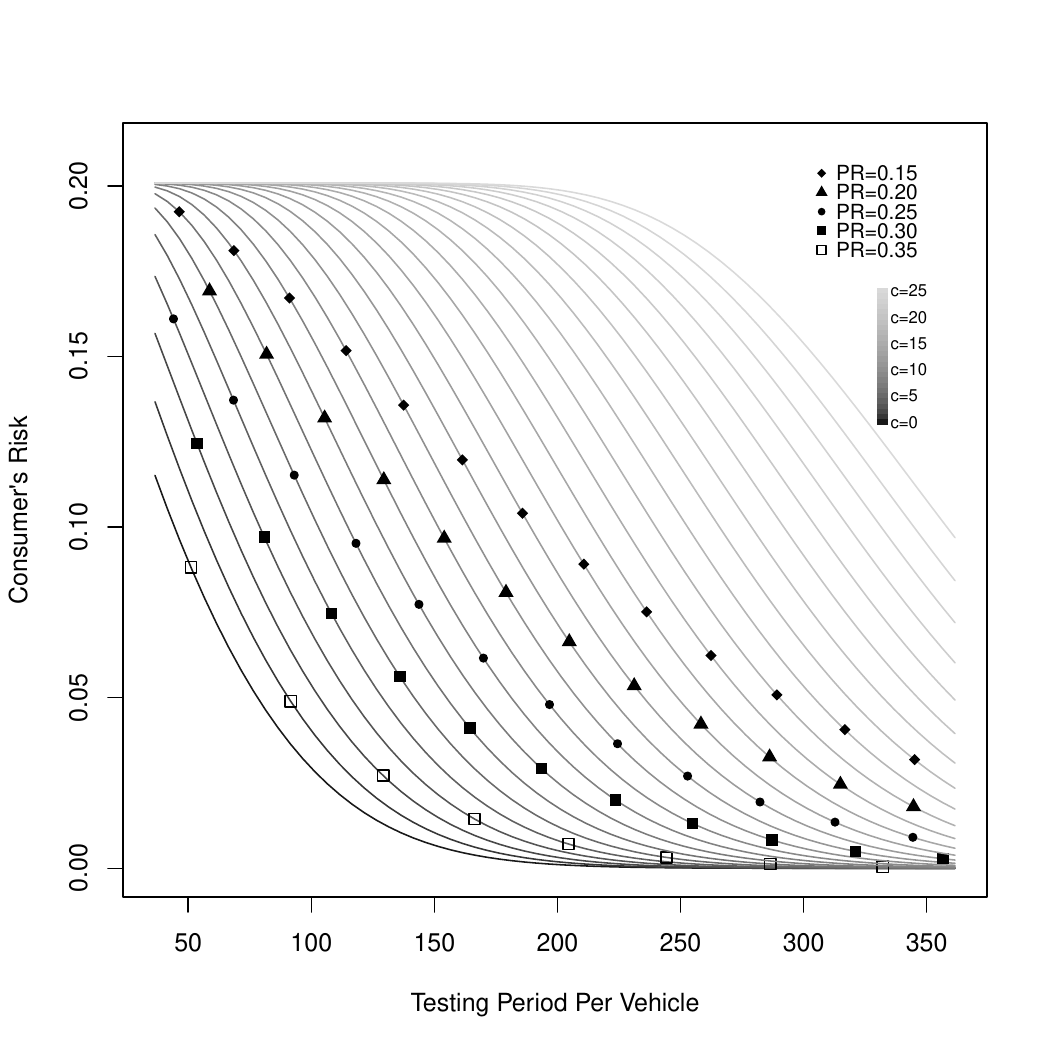} &
\includegraphics[width=0.50\textwidth]{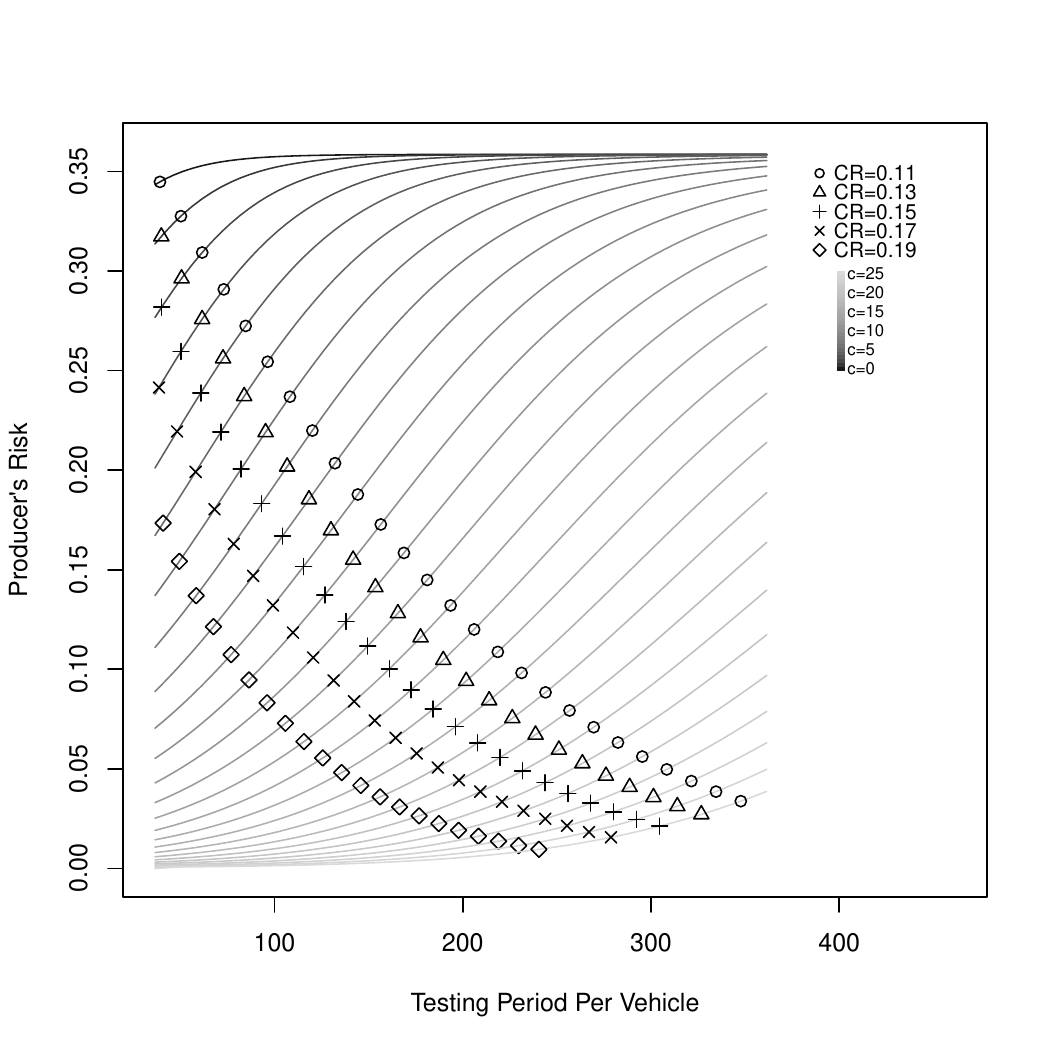}\\
(c) $\taut$ vs. CR   & (d) $\taut$ vs. PR
\end{tabular}
\caption{All possible test plans for the multiple vehicles test under the NHPP. The plots presented above illustrate the interrelations between CR, PR, AP, and $\taut$. In each plot, test plans with the same $c$ values are represented on a curve marked by varying shades of gray. These shades progress from darker to lighter, indicating ascending $c$ values within the $[0,25]$ range. Each symbol indicates selected representative symbols for other criteria. }
\label{fig:all.possible.tests.nhpp.multiple}
\end{center}
\end{figure}

Under the NHPP framework, for each test plan $(\nt, \taut,c)$, all criteria values are calculated based on \eqref{eqn:pcr.nhpp} to \eqref{eqn:ap.nhpp}. Figure~\ref{fig:all.possible.tests.nhpp.single} shows the performance of all test plans within the examined range, based on the four criteria outlined in Section~\ref{subsec:NHPP.risk}, for testing a single vehicle. While Figure~\ref{fig:all.possible.tests.nhpp.multiple} shows the interrelationship between the four criteria across all the test plans for the fleet test scenario with $\nt=5$.

Although the Figures \ref{fig:all.possible.tests.nhpp.single} and \ref{fig:all.possible.tests.nhpp.multiple} show a lot of similarities between the single vehicle and the fleet testing, we highlight the differences between the two scenarios under the NHPP model. Figures~\ref{fig:all.possible.tests.nhpp.single}(a) and~\ref{fig:all.possible.tests.nhpp.multiple}(a) show the relationship between CR and PR under the two test scenarios. In the single-vehicle test, each CR and PR curve exhibits a convex pattern which indicates a less severe trade-off between the two risk criteria for all the $c$ values. However, in the fleet test, for larger $c$ values, CR and PR curves exhibit a concave pattern, indicating a more severe trade-off between the two criteria.

Figures~\ref{fig:all.possible.tests.nhpp.single}(b) and~\ref{fig:all.possible.tests.nhpp.multiple}(b) show the relationship between $\taut$ and AP between the two criteria with different $c$ values under the two distinct test scenarios. The primary difference between these two test scenarios is that, for fixed $c$ and $\taut$, the AP in fleet testing is significantly lower than that in the single vehicle testing. This indicates that with a given maximum allowable failures and test duration per vehicle, testing more vehicles decreases the chance to pass the test.

Figures~\ref{fig:all.possible.tests.nhpp.single}(c) and~\ref{fig:all.possible.tests.nhpp.multiple}(c) show the relationship between $\taut$ and CR with highlighted different levels of CR and PR. Again we can observe increased concavity for smaller $c$ values in the fleet test scenario. This suggests that when testing multiple vehicles, there is a more severe trade-off between $\taut$ and CR at smaller values of $c$ compared to testing a single vehicle.

Figures~\ref{fig:all.possible.tests.nhpp.single}(d) and~\ref{fig:all.possible.tests.nhpp.multiple}(d) show the relationship between $\taut$ and PR with different levels of $c$ values under the two different testing scenarios. The specific pattern between $\taut$ and PR under these two different scenarios is different. In the single vehicle test, the curves between $\taut$ and PR are relatively flat. At a fixed value of $c$, the PR can be improved at the expense of increasing CR at a relatively slow speed by reducing the $\taut$. However, when testing multiple vehicles, at a fixed $c$ value, reducing the same amount of $\taut$ leads to a more substantial decrease in PR with the cost of increasing CR. This means that in the fleet test, given a fixed $c$, reducing $\taut$ will result in a larger improvement in PR at the cost of increasing CR compared to the single-vehicle test.

\begin{figure}[ht!]
\begin{center}
\includegraphics[width=0.65\textwidth]{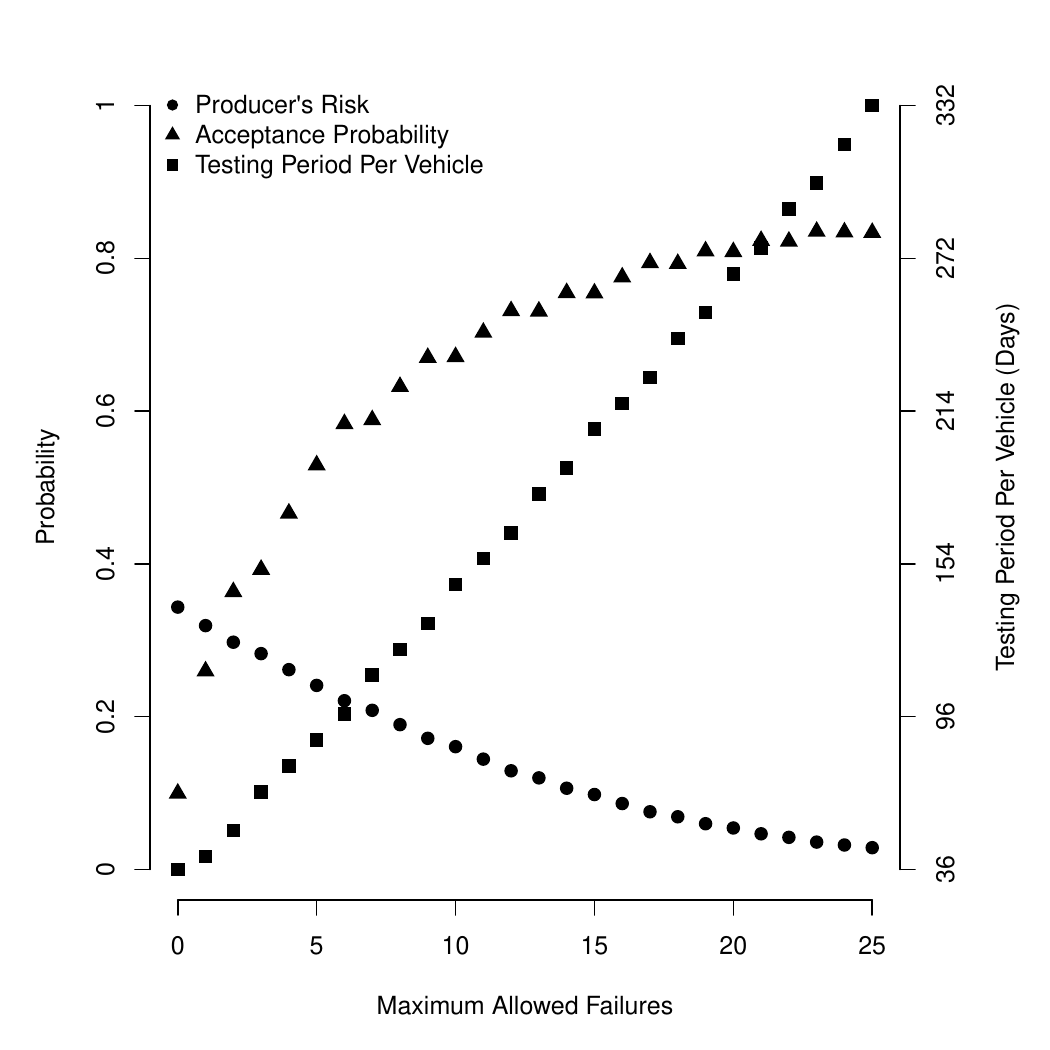}
\end{center}
\caption{The Pareto front of the optimal test plans under NHPP given the consumer's risk being controlled at or below 0.13. There are 26 test plans for the Pareto Front based on the PR, AP, and $\taut$, given by the constraint of CR does not surpass 0.13. Similar to what is depicted in HPP, the left axis denotes the scale for PR and AP, and the right axis for $\taut$. Each test plan choice corresponds to different values of $c$. }
\label{fig:PF.nhpp}
\end{figure}

To select a best potential test plan based on all possible test plans under the NHPP model, we focus on the fleet test vehicles scenario, since it mimics real-world AV test situations more closely. Then, we consider CR as the most important among the four criteria and prioritize the control of CR at or below $0.13$. Note that under the NHPP model, we set a relatively higher threshold for CR compared to that under the HPP model, this is because we anticipate that the CR may increase as the intensity varies throughout the testing period. Then, for all the test plans that meet the CR requirement, we remove inferior solutions by identifying the Pareto front with a set of non-dominated solutions based on the three other criteria. Figure~\ref{fig:PF.nhpp} shows the performance of all the test plans on the Pareto front based on PR, AP and $\taut$, given the constraint on CR.

From Figure~\ref{fig:PF.nhpp}, we can see that the Pareto front consists of 26 test plans with different values of $c$. Specifically, the Pareto front with the set of non-dominated solutions is organized from left to right, with $c$ values increasing from $0$ to $25$ on the $x$-axis. The left vertical axis, ranging from 0 to 1, corresponds to probability-related metrics. In contrast, the right vertical axis represents the total testing time $\taut$, with a range from $36$ to $332$ days. Similar to the Pareto front under the HPP model, we can see a significant trade-off between the three remaining criteria and the $c$ values under the NHPP model. However, at each $c$ value, there is a universal best plan based on simultaneously optimizing PR, AP, and $\taut$.

The final optimal testing plan can be selected based on different practitioner's priorities. If, for instance, the PR is considered the most important among the remaining criteria, say, the user is unwilling to accept a test plan with PR exceeding $0.15$, then the best test plan is to test $157$ days, and allow up to $11$ failures. This test plan will result in (1) CR at $0.126$, (2) PR at $0.144$, and (3) AP at $0.703$. However, if the user considers the AP as the most critical among the remaining criteria, particularly requiring an AP no less than $0.8$, then the best plan is to test for $252$ days with a maximum of $19$ failures. This will achieve (1) CR at $0.130$, (2) PR at $0.060$, and (3) AP at $0.810$. Alternatively, if the total testing period per vehicle is considered the most critical and a $\taut$ exceeding $200$ days per vehicle is unacceptable, then the best test plan will allow no more than $14$ failures for a successful test and results in (1) CR at $0.128$, (2) PR at $0.106$, and (3) AP at $0.755$.

%%%%%%%%%%%%%%%%%%%%%%%%%%%%%%%%%%%%%%%%%%%%%%%%%%%%%%%%%%%%%%%%%%%%%%%%%%%%%%%%%%%%%%%%%

%%%%%%%%%%%%%%%%%%%%%%%%%%%%%%%%%%%%%%%%%%%%%%%%%%%%%%%%%%%%%%%%%%%%%%%%%%%%%%%%%%%%%%%%%
\section{Conclusions and Areas for Future Research}\label{sec:conclusion}
%%%%%%%%%%%%%%%%%%%%%%%%%%%%%%%%%%%%%%%%%%%%%%%%%%%%%%%%%%%%%%%%%%%%%%%%%%%%%%%%%%%%%%%%%
This paper focuses on developing statistical methods for planning AV reliability assurance tests by using the recurrent disengagement events data from the CA self-driving program. We examine different aspects of the assurance test plans including (1) the consumer and producer's risks, (2) the probability of having a successful test, and (3) the total testing days or the testing days per vehicle. In addition, we thoroughly investigate the interrelationships among the four criteria under the HPP and NHPP models. We demonstrate that obtaining a deeper understanding of the interrelations between the test criteria, and how it is affected by assigned parameters can provide key insight into decisions related to assurance test planning in the field of AV testing and other related areas. Furthermore, understanding the trade-off between CR and PR can reshape the assurance test planning strategies, prompting them to weigh multiple dimensions and thus make the best choice for meeting their test objectives.

Another aspect of the analysis presented in this paper involves the use of the Pareto front approach to filter out inferior test plans. We then use the set of non-dominant solutions to guide the decision-making process, aligning with the priorities of practitioners and the objectives of the test. Specifically, in this paper, our primary focus is on controlling the CR, which is a common priority for many assurance tests. This strategy leads to a set of optimal solutions, each as a universally optimal plan that optimizes the three other criteria at each possible $c$ value. Given the identified set of superior solutions and a better understanding of the trade-offs, practitioners can make more informed decisions based on their available resources and the need to meet the planning goals.

For future work, we first plan to consider different event intensity functions such as the regression-type model in the form of $\lambda_{i} [t;\xvec_i(t),\thetavec] = \lambda_{0} (t;\thetavec)\exp[\beta x_{i}(t)].$
The analysis presented in this paper assumes a constant mileage effect for each test unit. However, in real-world scenarios, the mileage-driven function might vary for each test unit $i$. It would be interesting to use a regression-type model for the event intensity function to account variation in the milage driven by different vehicles. Other forms of $g(\cdot)$ can also be considered. For example, \shortciteN{shiau2010optimal} suggested using the exponential distribution to model the miles driven per day by drivers, resulting in the mileage effect function taking the following form: $g[x_{i}(t)]=x_{i}(t) \gamma \exp[-\gamma x_{i}(t)]$, for $\gamma>0$.

In addition, there are more aspects about the risk factors that might be of interest in the decision making. In our paper, we only consider CR, PR, AP, and the costs, which include the total testing period under the HPP model and the testing period per vehicle under the NHPP model. However, as indicated by \shortciteN{lu2016multiple}, we could consider a boarder aspect of the cost. For example, there is a potential additional cost related to CR due to the release of unacceptable products, resulting in increased warranty costs and loss of customers. On the other hand, the potential cost related to the PR is the unnecessary cost generated by rejecting a good product and hence requesting additional re-testing or re-design of the product. In future work for AV test planning, we plan to directly incorporate costs associated with CR and PR for assessing the performance of the test plans. We also plan to extend the historical period by incorporating more historical data for the disengagement events data and mileage information. By reaching out to the CA DMV, it is possible that we can access more historical data beyond the two-year period in the current study.

%\iffalse
%%%%%%%%%%%%%%%%%%%%%%%%%%%%%%%%%%%%%%%%%%%%%%%%%%%%%%%%%%%%%%%%%%%%%%%%%%%%%%%%%%%%%%%%%%%%%%%%%%%%%%%%%%%%%%%%%
\section*{Acknowledgments}
%%%%%%%%%%%%%%%%%%%%%%%%%%%%%%%%%%%%%%%%%%%%%%%%%%%%%%%%%%%%%%%%%%%%%%%%%%%%%%%%%%%%%%%%%%%%%%%%%%%%%%%%%%%%%%%%%
The authors acknowledge the Advanced Research Computing program at Virginia Tech for providing computational resources. The work by Hong was partially supported by the Virginia Tech College of Science Research Equipment Fund.

%\fi

%%%%%%%%%%%%%%%%%%%%%%%%%%%%%%%%%%%%%%%%%%%%%%%%%%%%%%%%%%%%%%%%%%%%%%%%%%%%%%%%%%%%%%%%%%%%%%%%%
%\bibliographystyle{chicago}
%\bibliography{ref}

\end{document}